\documentclass[aps,prb,twocolumn,showpacs,superscriptaddress,floatfix]{revtex4-1}
\usepackage{graphicx,graphics}
\usepackage{dcolumn}
\usepackage{amsmath,amssymb,amsfonts}
\usepackage{latexsym,verbatim}
\usepackage{bm}
\usepackage{color}
\usepackage{ulem}
\usepackage[breaklinks=true,colorlinks,citecolor=blue,linkcolor=blue,urlcolor=blue]{hyperref}

\def\be{\begin{equation}}
\def\ee{\end{equation}}

\begin{document}
\title{Plasmon losses due to electron-phonon scattering: the case of graphene encapsulated in hexagonal Boron Nitride}
\author{Alessandro Principi}
\email{principia@missouri.edu}
\affiliation{Department of Physics and Astronomy, University of Missouri, Columbia, Missouri 65211, USA}	
\author{Matteo Carrega}
\affiliation{NEST, Istituto Nanoscienze-CNR and Scuola Normale Superiore, I-56126 Pisa, Italy}
\affiliation{SPIN-CNR, Via Dodecaneso 33, 16146 Genova, Italy}
\author{Mark Lundeberg}
\affiliation{ICFO - Institut de Ci\`encies Fot\`oniques, Mediterranean Technology Park, Av. Carl Friedrich Gauss 3, E-08860 Castelldefels, Barcelona, Spain}
\author{Achim Woessner}
\affiliation{ICFO - Institut de Ci\`encies Fot\`oniques, Mediterranean Technology Park, Av. Carl Friedrich Gauss 3, E-08860 Castelldefels, Barcelona, Spain}
\author{Frank H.L. Koppens}
\affiliation{ICFO - Institut de Ci\`encies Fot\`oniques, Mediterranean Technology Park, Av. Carl Friedrich Gauss 3, E-08860 Castelldefels, Barcelona, Spain}
\author{Giovanni Vignale}
\affiliation{Department of Physics and Astronomy, University of Missouri, Columbia, Missouri 65211, USA}
\author{Marco Polini}
\affiliation{NEST, Istituto Nanoscienze-CNR and Scuola Normale Superiore, I-56126 Pisa, Italy}
\affiliation{Istituto Italiano di Tecnologia, Graphene Labs, Via Morego 30, I-16163 Genova, Italy}
\begin{abstract}
Graphene sheets encapsulated between hexagonal Boron Nitride (hBN) slabs display superb electronic properties due to very limited scattering from extrinsic disorder sources such as Coulomb impurities and corrugations. Such samples are therefore expected to be ideal platforms for highly-tunable low-loss plasmonics in a wide spectral range. In this Article we present a theory of collective electron density oscillations in a graphene sheet encapsulated between two hBN semi-infinite slabs (hBN/G/hBN). Graphene plasmons hybridize with hBN optical phonons forming hybrid plasmon-phonon (HPP) modes. We focus on scattering of these modes against graphene's acoustic phonons and hBN optical phonons, two sources of scattering that are expected to play a key role in hBN/G/hBN stacks. We find that at room temperature the scattering against graphene's acoustic phonons is the dominant limiting factor for hBN/G/hBN stacks, yielding theoretical inverse damping ratios of hybrid plasmon-phonon modes of the order of $50$-$60$, with a weak dependence on carrier density and a strong dependence on illumination frequency. We confirm that the plasmon lifetime is not directly correlated with the mobility: in fact, it can be anti-correlated.
\end{abstract}
\maketitle

\section{Introduction}
Hexagonal Boron Nitride (hBN), a wide-bandgap insulator, has recently emerged as a sort of ``magic'' substrate for exfoliated graphene sheets.  Early on, it was demonstrated~\cite{dean_naturenano_2010} that hBN is a much better substrate than ${\rm SiO}_2$---the ordinary substrate~\cite{andre_naturemater_2007} for much of the early work in graphene physics and devices---because its surface is much flatter and because it presents a much smaller number of charged impurities. Exfoliated graphene sheets deposited on hBN (G/hBN) or graphene sheets that are encapsulated in hBN (hBN/G/hBN) have therefore much larger mobilities~\cite{dean_naturenano_2010} than those that are deposited on ${\rm SiO}_2$.  Subsequently, vertical stacks~\cite{verticalheterostructures} comprising graphene and hBN have been used for proof-of-concept devices such as field-effect tunneling transistors~\cite{britnell_science_2012} and fundamental studies of electron-electron interactions~\cite{elias_naturephys_2011,yu_pnas_2013,gorbachev_naturephys_2012}. 
More recent experimental work~\cite{yankowitz_naturephys_2012,ponomarenko_nature_2013,dean_nature_2013,hunt_science_2013} has demonstrated that hBN substantially alters the electronic spectrum of the massless Dirac fermion (MDF)~\cite{castroneto_rmp_2009} carriers hosted in a nearby graphene sheet. Indeed, when graphene is deposited on hBN, it displays a moir\'e pattern~\cite{xue_naturemater_2011,decker_nanolett_2011}, a modified tunneling density of states~\cite{yankowitz_naturephys_2012}, and self-similar transport characteristics in a magnetic field~\cite{ponomarenko_nature_2013,dean_nature_2013,hunt_science_2013}. This spectral reconstruction of the MDF energy-momentum dispersion relation is, however, only relevant in the case of long-wavelength moir\'e superlattices (superlattice period $\gtrsim 10~{\rm nm}$), which occur when the twist angle between the graphene and hBN crystals is small. Short-wavelength superlattices yield changes of the MDF spectrum at dopings that are not achievable by electrostatic gating.

Finally, we would like to mention that the authors of Ref.~\onlinecite{wang_science_2013} have demonstrated that hBN/G/hBN samples, in which the role of contact resistance is minimized by using a suitable geometry, can display very large mobilities, which are solely limited by scattering of electrons against graphene's acoustic phonons. According to Boltzmann-transport theory~\cite{dassarma_rmp_2011}, this scattering mechanism yields~\cite{Hwang_prb_2008} a mobility $\mu$ that decreases like $\sim 1/n$ with increasing carrier density $n$---in good agreement with experimental data~\cite{wang_science_2013}-- and, therefore, a Drude dc transport scattering time $\tau_{\rm tr}$ that decreases like $1/\sqrt{n}$ in the same limit. 
We remind the reader that, in the same theoretical framework and by virtue of screening, scattering against charged impurities yields a mobility that increases with increasing carrier density~\cite{dassarma_rmp_2011}.

High-quality vertical heterostructures comprising graphene and hBN crystals may have a large impact on the success of graphene plasmonics~\cite{grapheneplasmonics}, an emerging field of research that has recently attracted a great deal of attention. The goal of graphene plasmonics is to exploit the interaction of infrared light with ``Dirac plasmons" (DPs)---the self-sustained density oscillations of the MDF liquid in a doped graphene sheet~\cite{Diracplasmons}---for a variety of applications such as infrared~\cite{freitag_naturecommun_2013} and Terahertz~\cite{vicarelli_naturemater_2012} photodetectors, strong light-matter interactions~\cite{Koppens_nanolett_2011}, enhanced light absorption~\cite{Thongrattanasiri_prl_2012} and bio-sensing~\cite{Kravets_naturemater_2013,Li_nanolett_2014}. Interest in graphene plasmonics considerably increased after two experimental groups~\cite{fei_nature_2012,chen_nature_2012} showed that the DP wavelength is much smaller than the illumination wavelength, allowing an extreme concentration of electromagnetic energy, and that it is easily gate tunable. 

These early experiments, based on scattering-type near-field optical spectroscopy (s-SNOM), were not optimized to minimize DP losses and therefore maximize the plasmon inverse damping ratio. Microscopic calculations targeting the role of electron-electron interaction effects~\cite{Principi_prb_2013_1} beyond the random phase approximation (RPA) and charged impurity scattering~\cite{Principi_prb_2013_2} indicate that losses can be strongly reduced by using hBN rather than ${\rm SiO}_2$ as a substrate. Indeed hBN has both a larger static dielectric constant, thus suppressing the strength of electron-electron interactions, and a much lower impurity concentration than ${\rm SiO_2}$.
The impact of electron-phonon scattering on the lifetime of HPP modes was recently addressed in Ref.~\onlinecite{yan_naturephoton_2013}, where the authors showed that the inverse damping ratio of mid-infrared HPP modes is strongly limited by scattering against optical phonons of both the SiO$_2$ substrate and the graphene sheet. The HPP mode damping rate was  estimated by introducing a self-energy correction in the {\it local} conductivity $\sigma(q,\omega)\simeq \sigma(\omega)$, which took into account electron-impurity, electron-phonon, and edge scattering. The contribution of electron-phonon interactions to the HPP lifetime was assumed to be independent of momentum. The dependence of the damping rate on momentum stemmed from the contribution due to the scattering against the edges of the sample.

In this Article we present a theoretical study of the performance of hBN/G/hBN stacks for applications in the field of graphene plasmonics. More precisely, we present a microscopic theory of the damping rate of HPP modes in a graphene sheet encapsulated between two hBN semi-infinite slabs. We focus on scattering of HPP modes in a graphene sheet against i) graphene's acoustic phonons and ii) hBN optical phonons, two sources of scattering that are expected to play a key role in limiting the lifetime of collective density oscillations in hBN/G/hBN stacks. Besides achieving good quantitative agreement with recent experimental work~\cite{Koppens}, we confirm the important fact that the plasmon lifetime is not necessarily correlated with the mobility~\cite{Principi_prb_2013_1,Principi_prb_2013_2} (i.e.~with the transport lifetime that controls the uniform dc conductivity). More accurately, the plasmon lifetime is controlled by the non-local conductivity $\sigma(q,\omega_{\rm p}(q))$, where $q$ is the wavevector and $\omega_{\rm p}(q)$ is the plasmon frequency. In the present case, retaining both the wavevector- and frequency-dependence of the non-local conductivity, we are able to show that the plasmon lifetime is in fact {\it anti-correlated} with the transport mobility. Indeed, while the mobility decreases with increasing carrier density~\cite{Hwang_prb_2008}, the plasmon lifetime shows exactly the opposite behavior.

This Article is organized as follows. In Sect.~\ref{sect:qualityfactor} we introduce the HPP mode inverse damping ratio $Q$ and relate it to the microscopic {\it non-local} dynamical conductivity $\sigma(q,\omega)$ of a 2D electron liquid embedded in a medium with a generic frequency-dependent dielectric function $\varepsilon_{\rm s}(\omega)$. In Sect.~\ref{sect:theory} we present our microscopic theory of the HPP mode dispersion relation and losses in hBN/G/hBN stacks due to electron-phonon scattering. In Sects.~\ref{sect:intrinsic_acoustic} and~\ref{sect:optical_substrate} we describe the details of the electron-phonon interactions we have considered, i.e. scattering of 2D MDFs against graphene's acoustic phonons and hBN optical phonons, respectively. Finally, in Sect.~\ref{sect:summary} we report a summary of our main results and conclusions. We have presented a number of relevant technical details in four Appendices, with the aim of making our Article as self-contained as possible.

\section{Weak-scattering theory of the inverse damping ratio}
\label{sect:qualityfactor}
In this Section we briefly derive general expressions for the inverse damping ratio $Q$ of a self-sustained oscillation in the density channel of an electron liquid~\cite{Principi_prb_2013_1,Principi_prb_2013_2}. 

To connect our theoretical results to the observables in s-SNOM experiments~\cite{chen_nature_2012,Gerber_prl_2014}, we assume the mode frequency to be a purely real quantity, which is fixed by the illumination frequency $\omega$. The mode damping rate $\gamma_{\rm p}$ is encoded in the imaginary part of the {\it complex} collective mode momentum $q_{\rm p} \equiv q_1 + i q_2$. More precisely, we define $\gamma_{\rm p}$ as
\begin{equation}\label{eq:damping_rate}
\gamma_{\rm p} \equiv \frac{q_2}{q_1}~.
\end{equation}
The inverse damping ratio $Q$ is the inverse of the damping rate, i.e.~$Q = \gamma^{-1}_{\rm p}$.

On the general grounds of linear response theory~\cite{Giuliani_and_Vignale,Pines_and_Nozieres}, the dispersion of HPP modes is calculated by solving the equation
\begin{equation} \label{eq:collective_modes_definition}
1 - V(q_{\rm p},\omega) {\widetilde \chi}_{nn}(q_{\rm p},\omega) = 0~,
\end{equation}
where
\begin{equation}\label{eq:dressedinteraction}
V(q,\omega) = \frac{v_{\bm q}}{\varepsilon_{\rm s}(\omega)}
\end{equation}
is an effective electron-electron interaction screened by a suitable substrate dielectric function $\varepsilon_{\rm s}(\omega)$ and ${\widetilde \chi}_{nn}(q_{\rm p},\omega)$ is the proper density-density linear response function~\cite{Giuliani_and_Vignale}. In Eq.~(\ref{eq:dressedinteraction}) $v_{\bm q} = 2\pi e^2/q$ is the 2D Fourier transform of the bare Coulomb interaction. The precise functional dependence of $\varepsilon_{\rm s}(\omega)$ on the illumination frequency $\omega$ is not specified in this Section. We stress that $q_{\rm p}\equiv q_{\rm p}(\omega)$ is defined as the solution of Eq.~(\ref{eq:collective_modes_definition}) at fixed illumination frequency $\omega$, while $q$ is hereafter a generic wavevector.

The causal (i.e.~retarded) density-density response function $\chi_{nn}(q,\omega)$ can be expressed in terms of the non-local frequency-dependent conductivity $\sigma(q,\omega)$ as follows~\cite{Giuliani_and_Vignale,Pines_and_Nozieres}
\begin{equation} \label{eq:chi_nn_sigma_relation}
\chi_{nn}(q,\omega) = \frac{q^2}{i e^2 \omega} \sigma(q,\omega)~.
\end{equation}
In the long-wavelenght $q/k_{\rm F} \ll 1$ limit and in two spatial dimensions~\cite{Giuliani_and_Vignale}, ${\widetilde \chi}_{nn}(q\ll k_{\rm F},\omega) = \chi_{nn}(q\ll k_{\rm F},\omega)$. Here, $k_{\rm F}$ is the Fermi wave number.

Since we are interested in describing scattering of collective modes against weak disorder, we approximate the non-local conductivity in the following generalized Drude form~\cite{Principi_prb_2013_1,Principi_prb_2013_2}:
\begin{equation} \label{eq:sigma_def}
\sigma(q,\omega) \simeq \frac{{\cal D}_0/\pi}{-i\omega +1/\tau(q,\omega)}~.
\end{equation}
In the spirit of the RPA~\cite{Giuliani_and_Vignale,Pines_and_Nozieres}, ${\cal D}_0$ is the Drude weight of a system of non-interacting 2D MDFs, ${\cal D}_0 = 4 \varepsilon_{\rm F} \sigma_{\rm uni}$, with $\varepsilon_{\rm F} = \hbar v_{\rm F} k_{\rm F}$ the MDF Fermi energy, $\sigma_{\rm uni} = N_{\rm f} e^2/(16\hbar)$ the universal optical conductivity~\cite{kotov_rmp_2012}, and $N_{\rm f} =4$ the number of fermion flavors~\cite{kotov_rmp_2012} in graphene. Corrections beyond RPA to the Drude weight, stemming from the lack of Galilean invariance of the 2D MDF model, have been worked out in Refs.~\onlinecite{abedinpour_prb_2011,levitov_prb_2013} and will be neglected in this work for the sake of simplicity.

Using Eq.~(\ref{eq:chi_nn_sigma_relation}) in Eq.~(\ref{eq:collective_modes_definition}) and solving for $q_{\rm p}$ we get
\begin{equation} \label{eq:q_p_sigma_epsilon}
q_{\rm p} = \frac{i\varepsilon_{\rm s}(\omega) \omega}{2\pi \sigma(q_{\rm p},\omega)}~.
\end{equation}

In the limit $q_2\ll q_1$ and neglecting terms containing the product $\Im m[\varepsilon_{\rm s}(\omega)] \Re e[\sigma(q_1,\omega)]$, Eq.~(\ref{eq:q_p_sigma_epsilon}) allows us to write
\begin{equation} \label{eq:gamma_p_definition}
\gamma_{\rm p} = \frac{\Im m[\varepsilon_{\rm s}(\omega)]}{\Re e[\varepsilon_{\rm s}(\omega)]} +\frac{\Re e [\sigma(q_1,\omega)]}{\Im m [\sigma(q_1,\omega)]}~,
\end{equation}
or, equivalently,
\begin{equation} \label{eq:Q_definition}
Q = \frac{1}{\displaystyle \frac{\Im m[\varepsilon_{\rm s}(\omega)]}{\Re e[\varepsilon_{\rm s}(\omega)]} +\frac{\Re e [\sigma(q_1,\omega)]}{\Im m [\sigma(q_1,\omega)]}}~.
\end{equation}
Eq.~(\ref{eq:Q_definition}) is the most important result of this Section.

The first term on the right-hand side of Eq.~(\ref{eq:gamma_p_definition}),
\begin{equation}\label{eq:substrateonly}
r^{-1}(\omega) \equiv \frac{\Im m[\varepsilon_{\rm s}(\omega)]}{\Re e[\varepsilon_{\rm s}(\omega)]}~,
\end{equation}
encodes the contribution to the damping rate that is solely controlled by the dissipative component of the substrate dielectric function $\varepsilon_{\rm s}(\omega)$ evaluated at the illumination frequency. The processes that are responsible for this dissipation have nothing to do with the presence of the graphene layer.  On the other hand, the second term on the right-hand side of Eq.~(\ref{eq:gamma_p_definition}), can be considered as an {\it intrinsic} damping rate, which involves electronic processes in the graphene layer. Indeed, expanding the right-hand side of Eq.~(\ref{eq:sigma_def}) in the weak-scattering $\omega \tau(q,\omega) \gg 1$ limit, we immediately find
\begin{equation} \label{eq:ratio_R}
{\cal R}(q,\omega) \equiv \frac{\Re e[\sigma(q,\omega)]}{\Im m [\sigma(q,\omega)]} \simeq \frac{1}{\omega \tau(q,\omega)}
\end{equation}
and
\begin{equation} \label{eq:Q_final_expression}
Q \simeq \frac{r(\omega)}{\displaystyle 1 + \frac{r(\omega)}{\omega \tau(q,\omega)}}~.
\end{equation}

Any microscopic theory of the collective mode damping rate requires the calculation of the quantity $\tau(q,\omega)$. 
Combining Eqs.~(\ref{eq:chi_nn_sigma_relation}) and~(\ref{eq:sigma_def}) we find that, in the weak-disorder $\omega \tau(q,\omega) \gg 1$ limit,
\begin{equation} \label{eq:lifetime_def}
\frac{1}{\tau(q,\omega)} \simeq -\frac{\pi e^2 \omega^3}{{\cal D}_0 q^2} \Im m[\chi_{nn} (q,\omega)]~.
\end{equation}
This is a very convenient expression~\cite{Principi_prb_2013_1,Principi_prb_2013_2} that will be used below to calculate damping rates due to electron-phonon scattering.

In the remainder of this Article we set $\hbar = 1$.

\section{Microscopic theory of losses due to electron-phonon scattering}
\label{sect:theory}

\subsection{Model Hamiltonian}
We consider the following model Hamiltonian
\begin{equation}\label{eq:Hamiltonian}
{\hat {\cal H}} = {\hat {\cal H}}_0 + {\hat {\cal H}}_{\rm ee} + {\hat {\cal H}}_{\rm ph} + {\hat {\cal H}}_{\rm e-ph}~.
\end{equation}
The first term in Eq.~(\ref{eq:Hamiltonian}) describes $\pi$-electrons in graphene at the level of a one-orbital tight-binding (TB) model. To keep the model as simple as possible, we set to zero all the hopping parameters but the nearest-neighbor one. The low-energy MDF limit will be taken only at the very end of the calculation, {\it after} carrying out all the necessary algebraic manipulations. As extensively discussed in Refs.~\onlinecite{Principi_prb_2013_1,Principi_prb_2013_2}, this procedure allows us to avoid problems associated with the introduction of a rigid ultraviolet cut-off, which breaks gauge invariance~\cite{abedinpour_prb_2011,levitov_prb_2013}. The non-interacting TB Hamiltonian reads
\begin{eqnarray} \label{eq:SM_non_int_H}
{\hat {\cal H}}_0 = \sum_{{\bm k} \in {\rm BZ}, \alpha, \beta} {\hat \psi}^\dagger_{{\bm k},\alpha} ({\bm f}_{{\bm k}} \cdot {\bm \sigma}_{\alpha\beta}) {\hat \psi}_{{\bm k},\beta}
~,
\end{eqnarray}
where the field operator ${\hat \psi}^\dagger_{{\bm k},\alpha}$ (${\hat \psi}_{{\bm k},\alpha}$) creates (annihilates) an electron with Bloch momentum ${\bm k}$, belonging to the sublattice~\cite{castroneto_rmp_2009} $\alpha = A,B$. The quantity ${\bm f}_{\bm k}$ is defined as~\cite{castroneto_rmp_2009}
\begin{eqnarray} \label{eq:SM_f_vec}
{\bm f}_{{\bm k}} = -t \sum_{i=1}^3 \left(\Re e\left[e^{-i {\bm k}\cdot{\bm \delta}_i}\right], -\Im m\left[e^{-i {\bm k}\cdot{\bm \delta}_i}\right]\right)
~.
\end{eqnarray}
Here $t\sim 2.8~{\rm eV}$ is the nearest-neighbor tunneling amplitude, while ${\bm \delta}_i$ ($i=1,\ldots,3$) are the vectors which connect an atom to its three nearest neighbors, {\it i.e.} ${\bm \delta}_1 = a \sqrt{3} {\hat {\bm x}}/2 + a {\hat {\bm y}}/2$, ${\bm \delta}_2 = -a \sqrt{3} {\hat {\bm x}}/2 + a{\hat {\bm y}}/2$, and ${\bm \delta}_3 = - a{\hat {\bm y}}$. Here $a \sim 1.42$~\AA~ is the Carbon-Carbon distance in graphene. The sum over ${\bm k}$ in Eq.~(\ref{eq:SM_non_int_H}) is restricted to the first Brillouin zone (BZ) and the Pauli matrices $\sigma^i_{\alpha\beta}$ ($i=x,y,z$) operate on the sublattice degrees of freedom. 

The TB problem posed by the Hamiltonian~(\ref{eq:SM_non_int_H}) can be easily solved analytically~\cite{castroneto_rmp_2009}. One finds the following eigenvalues $\varepsilon_{{\bm k},\lambda} = \lambda |{\bm f}_{\bm k}|$, with $\lambda=\pm$. These two bands touch at  two inequivalent points ($K$ and $K'$) in the hexagonal BZ. The low-energy MDF model is obtained from Eq.~(\ref{eq:SM_non_int_H}) by taking the limit $a\to 0$, while keeping the product $t a$ constant. In this limit ${\bm f}_{{\bm K}+{\bm k}} \to v_{\rm F} {\bm k}$, where $v_{\rm F} = 3 t a/2 \sim 10^6 ~{\rm m/s}$ is the density-independent Fermi velocity. It turns out to be more convenient to work in an eigenstate representation, in which the TB Hamiltonian reads 
\begin{equation}\label{eq:TB_eigenstate}
{\hat {\cal H}}_0 = \sum_{{\bm k}, \lambda} \varepsilon_{{\bm k},\lambda} {\hat c}^\dagger_{{\bm k},\lambda} {\hat c}_{{\bm k},\lambda}~,
\end{equation}
where ${\hat c}^\dagger_{{\bm k},\lambda}$ (${\hat c}_{{\bm k},\lambda}$) creates (annihilates) an electron in the single-particle eigenstate with eigenvalue $\varepsilon_{{\bm k},\lambda} = \lambda |{\bm f}_{\bm k}|$, with $\lambda=\pm$.

The second term in Eq.~(\ref{eq:Hamiltonian}) represents long-range Coulomb interactions between electrons. 
In the same representation as in Eq.~(\ref{eq:TB_eigenstate}), the Coulomb Hamiltonian reads~\cite{Giuliani_and_Vignale}
\begin{eqnarray} \label{eq:SM_interaction_H}
{\hat {\cal H}}_{\rm ee} = \frac{1}{2} \sum_{{\bm q}} v_{\bm q} {\hat n}_{\bm q} {\hat n}_{-{\bm q}}
~,
\end{eqnarray}
where the density operator is
\begin{eqnarray} \label{eq:SM_density_op}
{\hat n}_{\bm q} &=& \sum_{{\bm k},\lambda,\lambda'} {\cal D}_{\lambda\lambda'}({\bm k}-{\bm q}/2, {\bm k}+{\bm q}/2) {\hat c}^\dagger_{{\bm k}-{\bm q}/2,\lambda} {\hat c}_{{\bm k}+{\bm q}/2,\lambda'}
~,
\nonumber\\
\end{eqnarray}
and $v_{\bm q}$ is the 2D {\it discrete} Fourier transform of the real-space Coulomb interaction, which is a periodic function of the reciprocal-lattice vectors. Finally, in Eq.~(\ref{eq:SM_density_op}) we have introduced the ``density vertex"
\begin{eqnarray} \label{eq:SM_D_element}
{\cal D}_{\lambda\lambda'} ({\bm k}, {\bm k}') =
\frac{e^{i(\theta_{\bm k}-\theta_{{\bm k}'})/2} + \lambda\lambda' e^{-i(\theta_{\bm k}-\theta_{{\bm k}'})/2}}{2}
\end{eqnarray}
with $\theta_{\bm k} = {\rm Arg}[f_{{\bm k},x}+ i f_{{\bm k},y}]$. Here $\{f_{{\bm k}, i}, i= x,y\}$ denotes the Cartesian components of the vector ${\bm f}_{\bm k}$. In the low-energy MDF limit, $\theta_{{\bm K}+{\bm k}} \to \varphi_{\bm k}$, where $\varphi_{\bm k}$ is the angle between ${\bm k}$ and the ${\hat {\bm x}}$ axis.

Note that in writing Eq.~(\ref{eq:SM_interaction_H}) we have neglected the one-body operator proportional to the total number of particles, which avoids self-interactions~\cite{Giuliani_and_Vignale}, since it has no effect on the calculations that we will carry out below. The lifetime of collective modes outside the single particle-hole continuum is indeed determined by two-particle excitations only, which are generated by {\it two-body} operators.

Finally, the third and fourth term in Eq.~(\ref{eq:Hamiltonian}) represent the bare phonon Hamiltonian and the electron-phonon interaction Hamiltonian, respectively. These are given by
\begin{eqnarray} \label{eq:H_ph}
{\hat {\cal H}}_{\rm ph} = \sum_{{\bm q},\nu} \omega_{{\rm ph},\nu}({\bm q}) {\hat a}^\dagger_{{\bm q},\nu} {\hat a}_{{\bm q},\nu} ~,
\end{eqnarray}
and
\begin{eqnarray} \label{eq:H_e_ph}
{\hat {\cal H}}_{\rm e-ph} = \sum_{{\bm q},\nu} u_{{\bm q},\nu} {\hat n}_{{\bm q}_\perp} ({\hat a}_{-{\bm q},\nu} + {\hat a}^\dagger_{{\bm q},\nu})~.
\end{eqnarray}
Here ${\hat a}^\dagger_{{\bm q},\nu}$ (${\hat a}_{{\bm q},\nu}$) is the creation (annihilation) operator of a phonon belonging to the branch $\nu$, with momentum ${\bm q}$ measured from the BZ center, and energy $\omega_{{\rm ph},\nu}({\bm q})$. Here $\omega_{{\rm ph},\nu}(-{\bm q}) = \omega_{{\rm ph},\nu}({\bm q})$. Note that Eq.~(\ref{eq:H_ph}) is completely general and can be used to describe either graphene's intrinsic phonons, which are bound to the 2D graphene plane, or 3D phonons traveling in hBN. In the former (latter) case the phonon momentum ${\bm q}$ is a 2D (3D) vector. In the 3D case we write ${\bm q}\equiv ({\bm q}_\perp, q_\parallel)$, where ${\bm q}_\perp$ represents the projection of the 3D vector ${\bm q}$ on the graphene plane, while $q_\parallel$ is the component of ${\bm q}$ perpendicular to it.

In Eq.~(\ref{eq:H_e_ph}) we defined the electron-phonon interaction vertex $u_{{\bm q},\nu}$, which will be specified later in Sect.~\ref{sect:intrinsic_acoustic} and~\ref{sect:optical_substrate}. We stress again that also in Eq.~(\ref{eq:H_e_ph}) the vector ${\bm q}$ can be either 2D or 3D, according to the phonon modes of interest. In the case of optical phonons in the hBN substrate, only the component of the momentum along the graphene plane is conserved in the electron-phonon interaction.

In what follows we concentrate on a doped graphene sheet. For the sake of definiteness, we assume the system to be $n$-doped. As usual, results for a $p$-doped system can be easily obtained by appealing to the particle-hole symmetry of the model defined by Eq.~(\ref{eq:Hamiltonian}).

\subsection{HPP modes in hBN/G/hBN stacks}
\label{sect:collective_modes_hyperbolic}
As bulk graphite, hBN is a layered material: its response to electric fields is therefore highly anisotropic. Let ${\hat {\bm z}}$ be the axis perpendicular to the hBN planes, while ${\hat {\bm x}}$ and ${\hat {\bm y}}$ denote two orthogonal directions parallel to the hBN plane. The hBN dielectric tensor in this basis is diagonal and has the following uniaxial form~\cite{Geick_pr_1966},
\begin{equation}
{\hat \varepsilon}(\omega) = \left(
\begin{array}{ccc}
\varepsilon_x(\omega) & 0 & 0
\\
0 & \varepsilon_x(\omega) & 0
\\
0 & 0 & \varepsilon_z(\omega)
\end{array}
\right)
~.
\end{equation}
The dependence on frequency of the components $\varepsilon_x,\varepsilon_z$ of the dielectric tensor of bulk hBN is usually parametrized in the following form~\cite{Yu_and_Cardona}
\begin{eqnarray} \label{eq:BN_dielectric}
\varepsilon_\ell(\omega) &=& \epsilon_{\ell,\infty} + \frac{\epsilon_{\ell,0} - \epsilon_{\ell,\infty}}{1 - (\omega/\omega_{\ell}^{\rm T})^2 + i \gamma_\ell \omega/(\omega_{\ell}^{\rm T})^2}~,
\end{eqnarray}
with $\ell = z,x$. The parameters entering in Eq.~(\ref{eq:BN_dielectric}) are listed in Table~\ref{table_parameters} and have been taken from recent measurements~\cite{caldwell_arxiv_2014} on high-quality bulk hBN. Here $\epsilon_{\ell,0}$ and $\epsilon_{\ell, \infty}$ are the static and high-frequency dielectric constants, respectively, while $\omega^{\rm T}_{\ell}$ is the bulk transverse optical phonon frequency in the direction $\ell$. 
The bulk longitudinal optical phonon frequency $\omega^{\rm L}_{\ell}$ satisfies the Lyddane-Sachs-Teller (LST) relation~\cite{Yu_and_Cardona} $\omega^{\rm L}_{\ell} = \omega_{\ell}^{\rm T} \sqrt{\epsilon_{\ell,0}/\epsilon_{\ell,\infty}}$. 

\begin{table}[t]
\begin{ruledtabular}
\begin{tabular}{l | c c}
\, & $\ell=x$  & $\ell=z$	\\
\hline
$\epsilon_{\ell,0}$	& 6.41 & 3.0 \\
$\epsilon_{\ell,\infty}$ & 4.54 & 2.5 \\	
$\gamma_\ell~({\rm meV})$ & 0.82 & 0.23 \\
$\omega_{\ell}^{\rm T}~({\rm meV})$ & 168.0 & 94.2 \\
$\omega_{\ell}^{\rm L}~({\rm meV})$ & 199.6 & 103.2 \\
\end{tabular}
\end{ruledtabular}
\caption{The parameters entering the bulk hBN dielectric functions in Eq.~(\ref{eq:BN_dielectric}). These values have been extracted from Ref.~\onlinecite{caldwell_arxiv_2014}.\label{table_parameters}}
\end{table}

\begin{figure}[t]
\begin{center}
\begin{tabular}{c}
\includegraphics[width=0.99\columnwidth]{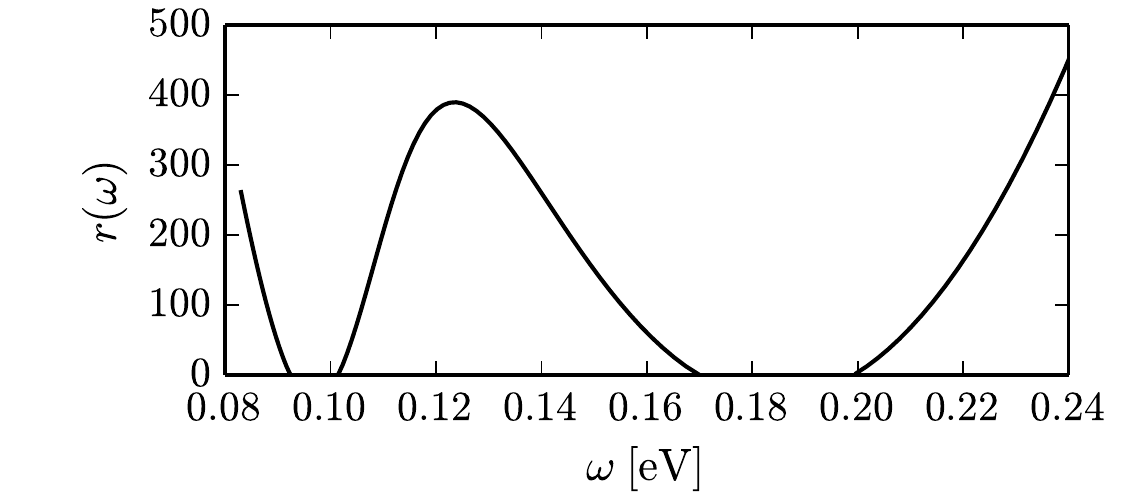}
\end{tabular}
\end{center}
\caption{The quantity $r(\omega)$ as defined in Eq.~(\ref{eq:substrateonly}) with $\varepsilon_{\rm s}(\omega)$ given in Eqs.~(\ref{eq:BN_dielectric})-(\ref{eq:effective-ee-interaction}). The parameters used in this plot are given in Tab.~\ref{table_parameters}.\label{fig:one}}
\end{figure}

\begin{figure}[t]
\begin{center}
\begin{tabular}{c}
\includegraphics[width=0.99\columnwidth]{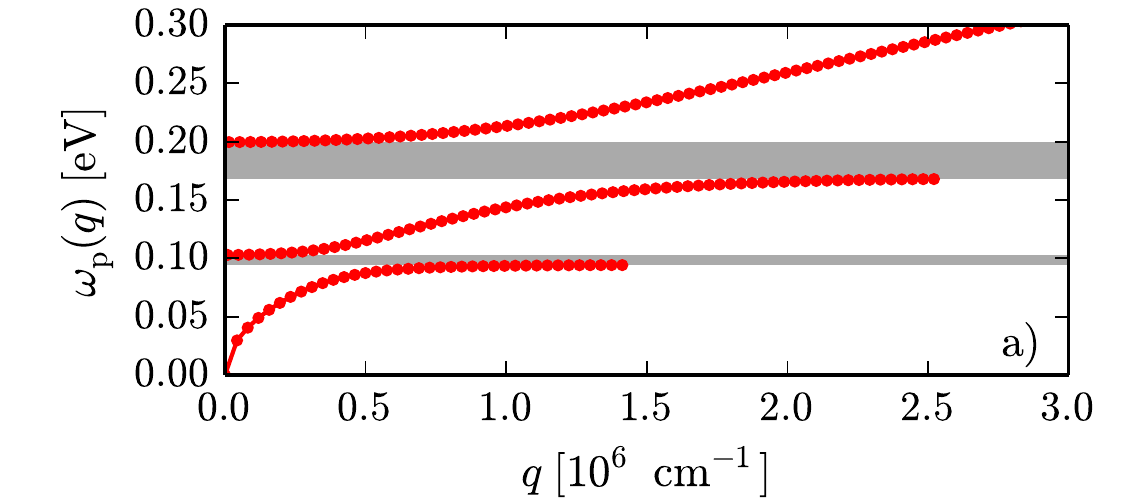}
\\
\includegraphics[width=0.99\columnwidth]{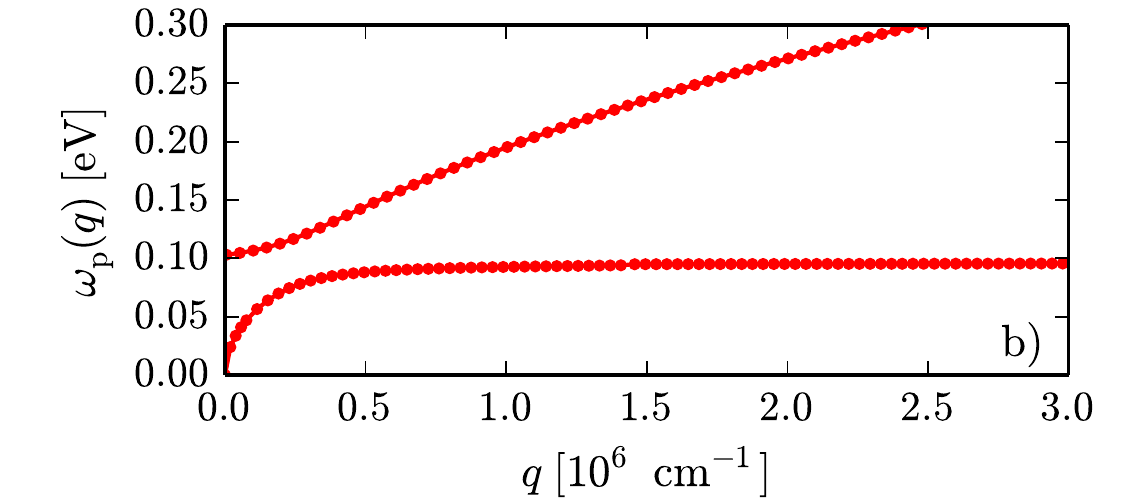}
\end{tabular}
\end{center}
\caption{(Color online) Panel a) The RPA dispersion relation $\omega_{\rm p}(q)$ of the HPP mode in hBN/G/hBN, as obtained from the solution of Eq.~(\ref{eq:RPA_dielectric_function}). Units are clearly indicated in the axes. Note the three branches of the HPP collective mode and the two reststrahlen bands (shaded regions).
Panel b) illustrates the same quantity in the limit in which one neglects the uniaxial anisotropy of hBN by forcing $\varepsilon_{x}(\omega) \to \varepsilon_{z}(\omega)$ in Eq.~(\ref{eq:effective-ee-interaction}) for all values of the illumination frequency $\omega$. In this case we see only two HPP branches.
\label{fig:two}}
\end{figure}

Electron-electron interactions in graphene are strongly modified when graphene is embedded between 
two half-spaces filled with hBN. As shown in Appendix~\ref{sect:ee_interaction_electrostatics}, the bare Coulomb potential $v_{\bm q} = 2\pi e^2/q$ is replaced by the dressed interaction in Eq.~(\ref{eq:dressedinteraction}) with
\begin{equation}\label{eq:effective-ee-interaction}
\varepsilon_{\rm s}(\omega) = \sqrt{\varepsilon_z(\omega)\varepsilon_x(\omega)}~.
\end{equation}
The resultant effective electron-electron interaction includes screening due to the hBN optical phonons only. We have not considered the renormalization of electron-electron interactions due to intrinsic acoustic phonons in graphene. Since the matrix element of the electron-acoustic phonon interaction vanishes in the long-wavelength $q \to 0$ limit (see Sect.~\ref{sect:intrinsic_acoustic}), intrinsic acoustic phonons in graphene do not affect the dispersion of the collective modes---see Fig.~\ref{fig:two}a)---in the limit $q \ll k_{\rm F}$ and $v_{\rm F} q \ll \omega \ll 2\varepsilon_{\rm F}$. In Fig.~\ref{fig:one} we plot the ratio $r(\omega)$ defined in Eq.~(\ref{eq:substrateonly}) with $\varepsilon_{\rm s}(\omega)$ given by Eq.~(\ref{eq:effective-ee-interaction}).

Self-sustained oscillations of the 2D MDF liquid in a graphene sheet embedded between two half-spaces filled with hBN can be found by solving the following RPA equation
\begin{equation}\label{eq:RPA_dielectric_function}
\varepsilon(q,\omega) \equiv 1 - V(q,\omega) \chi^{(0)}_{nn}(q,\omega) = 0~,
\end{equation}
where $V(q,\omega)$ is the effective electron-electron interaction in Eq.~(\ref{eq:dressedinteraction}) with $\varepsilon_{\rm s}(\omega)$ as in Eq.~(\ref{eq:effective-ee-interaction}) and $\chi^{(0)}_{nn}(q,\omega)$ is the well-known density-density response function of a non-interacting 2D MDF fluid~\cite{Diracplasmons}.

In Fig.~\ref{fig:two}a) we plot the solution of Eq.~(\ref{eq:RPA_dielectric_function}), as found by setting the hBN phonon damping rates $\gamma_{z}, \gamma_{x}$ to zero in Eq.~(\ref{eq:BN_dielectric}). We clearly see that hybridization between the ordinary DP mode~\cite{grapheneplasmonics,Diracplasmons} in a doped graphene sheet and hBN optical phonons yields {\it three} HPP branches~\cite{Amorim_prb_2012}. Furthermore, we note the presence of two forbidden regions, which are denoted by shaded areas in Fig.~\ref{fig:two}a). These correspond to the so-called hBN ``reststrahlen bands''.  The lower (upper) reststrahlen band is defined by the inequalities $\omega^{\rm T}_z < \omega < \omega^{\rm L}_z$ ($\omega^{\rm T}_x < \omega < \omega^{\rm L}_x$). In these spectral windows, the product $\varepsilon_x(\omega)\varepsilon_z(\omega)$ takes negative values and the substrate dielectric function (\ref{eq:effective-ee-interaction}) becomes imaginary. Indeed, as discussed earlier in the literature~\cite{caldwell_arxiv_2014,dai_science_2014}, hBN is a natural hyperbolic material~\cite{hyperbolicmaterials}. As shown in Appendix~\ref{sect:ee_interaction_electrostatics}, in the reststrahlen bands the HPP collective mode ``deconfines'' from the graphene sheet and propagates as a phonon inside bulk hBN. 

In the long-wavelength $q \to 0$ limit the lowest-energy branch behaves like $\omega_1(q \to 0) = \sqrt{N_{\rm f} e^2 \varepsilon_{\rm F} q/(2\epsilon_{z,0})}$, displaying a predominant 2D plasmon character. In the same limit the second branch $\omega_2(q)$ is gapped and behaves as 
$\omega_2(q\to 0) \to \omega^{\rm L}_z$, displaying a predominant longitudinal-phonon character. The two branches show an avoided crossing and switch their character in the short-wavelength $q \to \infty$ limit. The lower branch tends to a constant, $\omega_1(q\to \infty)  \to \omega^{\rm T}_z$, while $\omega_2(q \to \infty)$ follows the DP dispersion of an isolated graphene sheet. 

In Fig.~\ref{fig:two}b) we show what happens to the solution of Eq.~(\ref{eq:RPA_dielectric_function}) when the uniaxial anisotropy of hBN is neglected by forcing, for example, 
$\varepsilon_{x}(\omega) \to \varepsilon_{z}(\omega)$ at all frequencies. Showing results for an isotropic $\varepsilon_x = \varepsilon_z$ polar material allows us to make contact with earlier literature~\cite{Koch_prb_2010,Hwang_prb_2010} on different heterosystems such as graphene on ${\rm SiO}_2$ or ${\rm SiC}$. The necessary parameters for $\varepsilon_z(\omega)$ are listed in the last column of Table~\ref{table_parameters}. 
In this case, the effective dielectric function $\varepsilon_{\rm s}(\omega)$ is always real---since $\varepsilon_{z}(\omega) \varepsilon_{x}(\omega) = \varepsilon^2_z(\omega) > 0$---and the HPP mode displays only {\it two} branches as in earlier work on HPP modes in graphene on ${\rm SiO}_2$ or ${\rm SiC}$~\cite{Koch_prb_2010,Hwang_prb_2010}.

\subsection{Evaluation of the scattering rate due to electron-phonon interactions}
\label{sect:evaluation}

We now proceed to calculate $\tau(q,\omega)$ as from Eq.~(\ref{eq:lifetime_def}).

To this end, we evaluate $\chi_{nn}(q,\omega)$ on the right-hand side of Eq.~(\ref{eq:lifetime_def}) to {\it second} order in the strength of electron-phonon interactions. Following Refs.~\onlinecite{Principi_prb_2013_1,Principi_prb_2013_2}, we focus our attention on the longitudinal current-current response function $\chi_{\rm L}(q,\omega)$, which is related to $\chi_{nn}(q,\omega)$ by the following exact identity~\cite{Giuliani_and_Vignale}, 
\be \label{eq:continuity_equation}
\Im m[\chi_{nn}(q,\omega)] = \frac{q^2}{\omega^2}\Im m[\chi_{\rm L}(q,\omega)]~,
\ee
which holds for an isotropic, rotationally-invariant electron liquid~\cite{Giuliani_and_Vignale}. As explained in Refs.~\onlinecite{Principi_prb_2013_1,Principi_prb_2013_2}, Eq.~(\ref{eq:continuity_equation}) applies in our TB description after one takes the low-energy MDF limit. 

We then introduce~\cite{Principi_prb_2013_1,Principi_prb_2013_2} a unitary transformation generated by an Hermitian operator ${\hat F}$ that cancels exactly the electron-phonon interaction term from the Hamiltonian ${\hat {\cal H}}$ in Eq.~(\ref{eq:Hamiltonian}). For the sake of simplicity, this procedure is formally carried out by setting ${\hat {\cal H}}_{\rm ee} = 0$ in Eq.~(\ref{eq:Hamiltonian}): we will come back to the crucial role~\cite{Principi_prb_2013_2} played by electron-electron interactions below.

The Hermitian generator ${\hat F}$ is found by requiring that
$
{\hat {\cal H}}' = e^{i {\hat F}} {\hat {\cal H}} e^{-i {\hat F}} \equiv {\hat {\cal H}}_0 + {\hat {\cal H}}_{\rm ph}~,
$ 
i.e., the electron-phonon interaction is eliminated. This equation can be solved order by order in perturbation theory, by expanding ${\hat F} = \openone + {\hat F}_1 + {\hat F}_2 + \ldots$, where $\openone$ is the identity operator and ${\hat F}_n$ denotes the $n$-th order term in powers of the strength of electron-phonon interactions. We obtain a chain of equations connecting ${\hat F}_n$ to ${\hat {\cal H}}_{\rm e-ph}$. As an example, ${\hat F}_1$ can be easily found by solving the following equation: $[{\hat {\cal H}}_0 + {\hat {\cal H}}_{\rm ph}, i {\hat F}_1] = {\hat {\cal H}}_{\rm e-ph}$. 

Under the action of the Hermitian operator ${\hat F}$, the Hamiltonian ${\hat {\cal H}} \to {\hat {\cal H}}_0+{\hat {\cal H}}_{\rm ph}$ but relevant operators such as the density ${\hat n}_{\bm q}$ and the current ${\hat {\bm j}}_{\bm q}$ ones are dressed by electron-phonon interactions in a complicated manner, i.e.~${\hat n}_{\bm q} \to {\hat n}'_{\bm q} \equiv e^{i {\hat F}} {\hat n}_{\bm q} e^{-i {\hat F}}$ and 
${\hat {\bm j}}_{\bm q} \to {\hat {\bm j}}'_{\bm q} \equiv e^{i {\hat F}} {\hat {\bm j}}_{\bm q} e^{-i {\hat F}}$.
The dressed current operator, in particular, can be expanded in powers of the electron-phonon interaction as
$
{\bm q}\cdot{\hat {\bm j}}_{\bm q}' = {\bm q}\cdot{\hat {\bm j}}_{\bm q} + {\bm q}\cdot{\hat {\bm j}}_{1,{\bm q}} + {\bm q}\cdot{\hat {\bm j}}_{2,{\bm q}} + \ldots
$, 
where ${\hat {\bm j}}_{n,{\bm q}}$ is ${\cal O}(u_{\bm q})$. 

The zeroth-order contribution to ${\bm q}\cdot{\hat {\bm j}}_{\bm q}'$, {\it i.e.} ${\bm q}\cdot{\hat {\bm j}}_{{\bm q}}$, does {\it not} break momentum conservation by transferring part of the momentum ${\bm q}$ to the phonon subsystem. 
Indeed, ${\hat {\bm j}}_{\bm q}$ can only generate single particle-hole pairs with total momentum ${\bm q}$, which lie inside the particle-hole continuum. In turn, this implies that, in the limit $v_{\rm F} q \ll \omega \ll 2\varepsilon_{\rm F}$, the only non-vanishing second-order contribution in the strength of electron-phonon interactions to $\Im m[\chi_{\rm L}(q,\omega)]$ is given by $\Im m[\chi_{j_{1,x} j_{1,x}}(q{\hat {\bm x}},\omega)]$. We find
\begin{equation} \label{eq:SM_j_1_element}
{\bm q}\cdot{\hat {\bm j}}_{1,{\bm q}} = [i {\hat F}_1, {\bm q}\cdot{\hat {\bm j}}_{\bm q}] = {\cal A}_{\rm BZ}^{-1} \sum_{{\bm q}',\nu} u_{{\bm q}',\nu}
{\hat \Upsilon}_{{\bm q}, {\bm q}'_\perp} ({\hat a}_{-{\bm q}',\nu} + {\hat a}^\dagger_{{\bm q}',\nu})~,
\end{equation}
where ${\cal A}_{\rm BZ}$ is the BZ area.
 
It is clear from Eq.~(\ref{eq:SM_j_1_element}) that ${\bm q}\cdot{\hat {\bm j}}_{1,{\bm q}}$ breaks momentum conservation by transferring an amount ${\bm q}'_\perp$ to the phonon subsystem. In the limit $v_{\rm F} q \ll \omega \ll 2\varepsilon_{\rm F}$, the operator ${\hat \Upsilon}_{{\bm q}, {\bm q}'}$ reads
\begin{eqnarray} \label{eq:Upsilon_approx_0}
{\hat \Upsilon}_{{\bm q},{\bm q}'} &=&
\frac{v_{\rm F} q'_x}{k_{\rm F}\omega} {\hat \rho}_{{\bm q}+{\bm q}'}
- 
\Bigg[
2 \frac{v_{\rm F} q}{\omega^2} \frac{q'_x}{k_{\rm F}} \left(1 - \frac{q'^2}{4 k_{\rm F}^2} \right)
\nonumber\\
&-&
\frac{q'^2}{4 v_{\rm F} k_{\rm F}^3} 
\Bigg] {\hat j}_{{\bm q}',x}
~.
\end{eqnarray}
Note that the two terms on the right-hand side of Eq.~(\ref{eq:Upsilon_approx_0}) have the same physical dimensions, since the current operator scales with an extra power of the Fermi velocity with respect to the density operator.
The operator of Eq.~(\ref{eq:Upsilon_approx_0}) is suitable to calculate the imaginary part of the longitudinal current-current response function to second order in the strength of electron-phonon interactions. As we show in Appendix~\ref{app:response_function_finite_T}, this is given by the convolution of a non-interacting response function with the phonon propagator, and describes the decay of a quasiparticle of energy $\omega$ into a particle-hole pair with energy $\omega-\omega'$,  assisted by a phonon with energy $\omega'$. We use this information to further simplify Eq.~(\ref{eq:Upsilon_approx_0}). We note that the particle-hole pair is created by the one-body operator ${\hat \rho}_{{\bm q}+{\bm q}'}$, whose equation of motion reads $(\omega - \omega') {\hat \rho}_{{\bm q}+{\bm q}'} = -({\bm q}+{\bm q}')\cdot {\hat j}_{{\bm q}+{\bm q}', \alpha}$. This in turn implies that
\begin{eqnarray} \label{eq:Upsilon_approx}
{\hat \Upsilon}_{{\bm q},{\bm q}'} &=& 
- \sum_{\alpha=x,y} \left\{
\frac{v_{\rm F}}{\omega(\omega-\omega') k_{\rm F}} q'_x (q'_\alpha + q \delta_{\alpha,x}) 
\right.
\nonumber\\
&+&
\left.
2\left[ \frac{v_{\rm F}}{\omega^2} \frac{{\bm q}\cdot{\bm q}'}{k_{\rm F}} \left( 1 - \frac{q'^2}{4 k_{\rm F}^2} \right) - \frac{q'^2}{4 v_{\rm F} k_{\rm F}^3} \right] \delta_{\alpha,x}
\right\}
{\hat j}_{{\bm q}+{\bm q}', \alpha}
\nonumber\\
&\equiv&
- \sum_{\alpha=x,y} \Gamma^{({\rm ph})}_\alpha({\bm q},{\bm q}',\omega,\omega') {\hat j}_{{\bm q}+{\bm q}', \alpha}
~.
\end{eqnarray}
We stress that this equation is exact to linear order in the expansion in powers of $q/k_{\rm F}$.

The derivation of the imaginary part of the current-current response function to second order in the strength of electron-phonon interactions is carried out in Appendix~\ref{app:response_function_finite_T}. Here we report only the final result, which involves also taking the low-energy MDF limit.  We find
\begin{eqnarray} \label{eq:lifetime_final_T}
\frac{1}{\tau(q,\omega)} &=& - \frac{\pi e^2\omega}{{\cal D}_0} \sum_{\alpha,\beta,\nu} \int \frac{d^3{\bm q}'}{(2\pi)^3} \int_{-\infty}^{\infty} \frac{d\omega'}{\pi} u_{{\bm q}',\nu}^2  
\nonumber\\
&\times&
\big[n_{\rm B}(\omega') - n_{\rm B}(\omega'-\omega) \big]  \Gamma^{({\rm ph})}_\alpha({\bm q},{\bm q}'_\perp,\omega,\omega') 
\nonumber\\
&\times&
\Gamma^{({\rm ph})}_\beta({\bm q},{\bm q}'_\perp,\omega,\omega') \Im m~[{\cal D}^{({\rm ph})}_{\nu} (q',\omega')]
\nonumber\\
&\times&
\Im m~[\chi_{j_\alpha j_\beta}^{(0)} ({\bm q}+{\bm q}'_\perp,\omega-\omega')]~,
\end{eqnarray}
where $n_{\rm B}(\omega) = [\exp(\beta\omega)-1]^{-1}$ is the usual Bose-Einstein thermal factor with $\beta = (k_{\rm B} T)^{-1}$ and
\begin{eqnarray}\label{eq:phonon_propagator}
{\cal D}^{({\rm ph})}_{\nu}({\bm q},\omega) 
&=& \frac{2\omega_{{\rm ph}, \nu}({\bm q})}{(\omega - i \gamma_\nu)^2 - \omega_{{\rm ph}, \nu}^2({\bm q})}
\end{eqnarray}
is the phonon propagator~\cite{Mahan}. Eq.~(\ref{eq:lifetime_final_T}) is the most important result of this Section.

In Appendix~\ref{app:transport_time} we show that, in the limit $q=0$ and $\omega \to 0$, Eq.~(\ref{eq:lifetime_final_T}) reproduces the dc transport time $\tau_{\rm tr}$ for scattering of electrons against graphene's acoustic phonons as found e.g.~in Ref.~\onlinecite{Hwang_prb_2008}. 

We can now easily take into account electron-electron interactions, which were dropped at the beginning of Sect.~\ref{sect:evaluation}. This is done by replacing in Eq.~(\ref{eq:lifetime_final_T}) the longitudinal and transverse components of the non-interacting current-current response function $\chi_{j_\alpha j_\beta}^{(0)} ({\bm q},\omega)$ with the RPA current-current response $\chi_{j_\alpha j_\beta}^{({\rm RPA})} ({\bm q},\omega)$. We remind the reader (i) that the RPA dielectric function has been introduced in Eq.~(\ref{eq:RPA_dielectric_function}) and (ii) that the transverse RPA current-current response function coincides with the non-interacting one~\cite{Giuliani_and_Vignale}.

\section{Scattering of HPP modes against graphene's acoustic phonons}
\label{sect:intrinsic_acoustic}

We now consider the impact of graphene's intrinsic acoustic phonons on 
the damping rate of the HPP mode discussed in Sect.~\ref{sect:collective_modes_hyperbolic} and shown in Fig.~\ref{fig:two}a). In this case, ${\bm q}$ in Eq.~(\ref{eq:H_e_ph}) is a 2D wavevector in the graphene plane.

As far as the electron-acoustic phonon interaction vertex $u^{({\rm ac})}_{\bm q}$ is concerned, we have chosen  to follow earlier works~\cite{Hwang_prb_2008,dassarma_rmp_2011,Tse_prb_2009,Viljas_prb_2010} in which this is approximated in the following manner,
\begin{equation} \label{eq:acoustic_phonon_interaction}
|u_{\bm q}^{({\rm ac})}|^2 = \delta_{q_\parallel,0}\frac{{\widetilde D}^2 q^2}{2 \rho \omega_{\rm ph}(q)}~,
\end{equation}
where ${\widetilde D}$ is an effective deformation potential, $\rho=7.6 \times 10^{-8} ~{\rm g/cm}^2$ is the graphene's mass density, and $\omega_{\rm ph}(q) = {\widetilde v}_{\rm ph} q$ is an effective acoustic phonon dispersion. Here ${\widetilde v}_{\rm ph} \sim 0.02~v_{\rm F}\sim 20~{\rm km}/{\rm s}$~\cite{Hwang_prb_2008,dassarma_rmp_2011,Tse_prb_2009,Viljas_prb_2010}. The validity of the effective model (\ref{eq:acoustic_phonon_interaction}) has been recently confirmed by extensive first-principles calculations~\cite{Park_nanolett_2014,Sohier_arxiv_2014}. See, in particular, the discussion in Sect.~VIII of Ref.~\onlinecite{Sohier_arxiv_2014}.

In what follows, we determine the effective deformation potential ${\widetilde D}$ by requiring that the dc mobility
\begin{equation}\label{eq:mobility}
\mu = \frac{e \tau_{\rm tr}}{m_{\rm c}}~,
\end{equation}
with $m_{\rm c} = k_{\rm F}/v_{\rm F}$ the cyclotron mass and $\tau_{\rm tr} \equiv \lim_{\omega \to 0}\tau(0,\omega)$, matches the value measured in Ref.~\onlinecite{wang_science_2013} at $T =300~{\rm K}$. Following this procedure~\cite{footnote_fitting}, we obtain ${\widetilde D}=48.3~{\rm eV}$.

\begin{figure}[h!]
\begin{center}
\begin{tabular}{c}
\includegraphics[width=0.99\columnwidth]{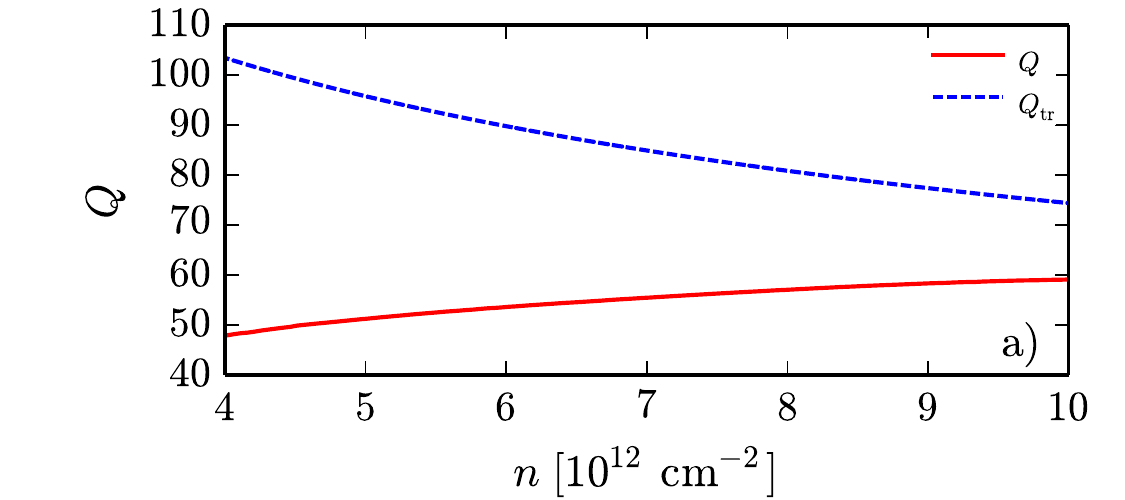}\\
\includegraphics[width=0.99\columnwidth]{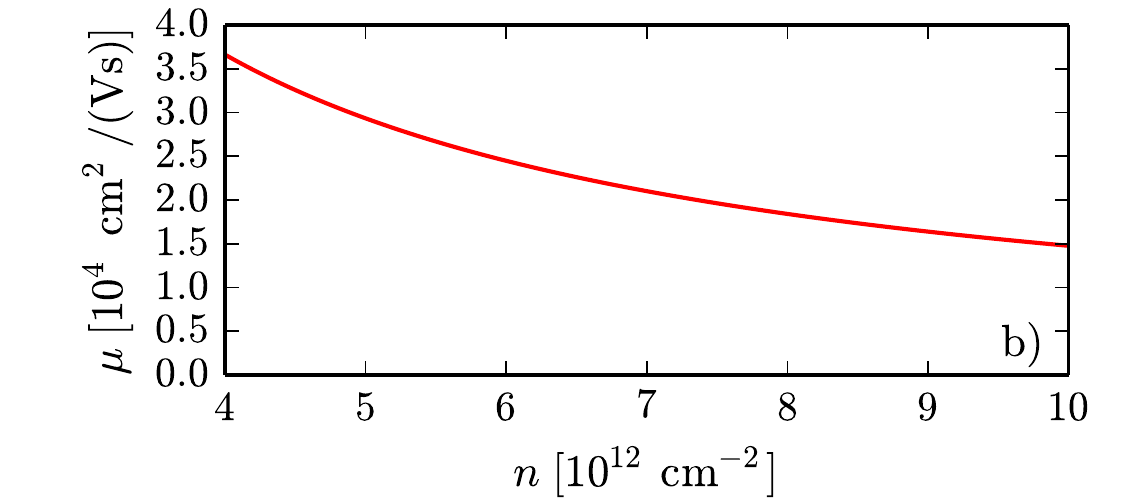}\\
\includegraphics[width=0.99\columnwidth]{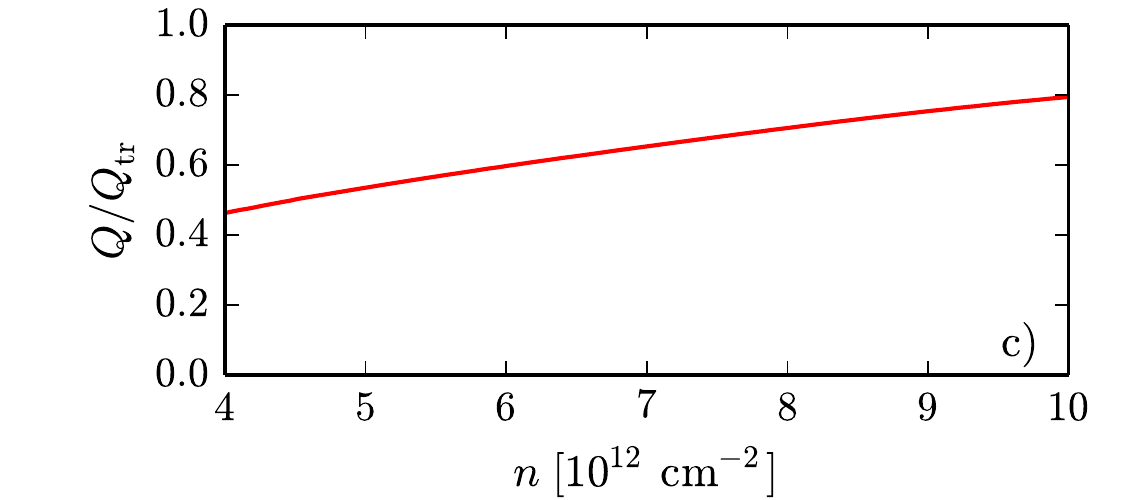}\\
\includegraphics[width=0.99\columnwidth]{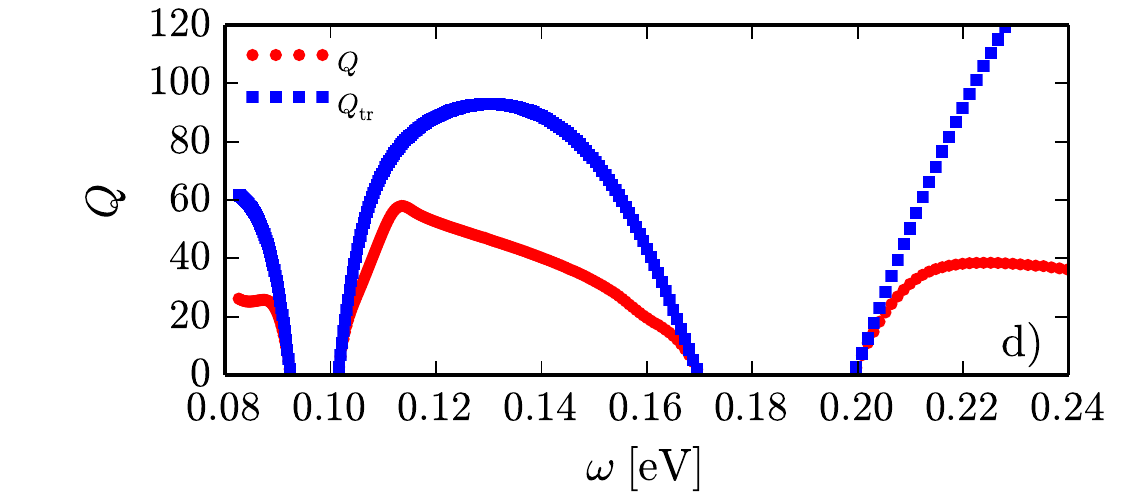}
\end{tabular}
\end{center}
\caption{(Color online) Panel a) The inverse damping ratio $Q$ of the HPP mode in a hBN/G/hBN stack. In this plot we show the impact of acoustic phonon scattering (solid line). The quantity $Q$ is plotted as a function of carrier density $n$ and for a fixed illumination frequency $\omega = 110~{\rm meV}$. As a comparison, we also plot the quantity $Q_{\rm tr}$, which is the inverse damping ratio $Q$ calculated from Eq.~(\ref{eq:Q_final_expression}) by replacing $\tau(q_1,\omega)$ with the dc transport scattering time $\tau_{\rm tr} = \lim_{\omega \to 0}\tau(0,\omega)$---dashed line. 
Panel b) The electron mobility in a hBN/G/hBN stack as a function of carrier density, as calculated from Eq.~(\ref{eq:mobility}) by considering scattering of electrons against graphene's acoustic phonons. Panel c) The ratio $Q/Q_{\rm tr}$, as extracted from panel a).
Panel d) Same as in panel a) but with $Q$ plotted as a function of the illumination frequency and for a fixed carrier density $n = 7.2\times 10^{12}~{\rm cm}^{-2}$. Note the two ``gaps'' due to the hBN reststrahlen bands. All data in this figure have been calculated at $T=300~{\rm K}$.\label{fig:three}}
\end{figure}

Our main results for the acoustic-phonon-scattering-limited inverse damping ratio $Q$ of the HPP mode shown in Fig.~\ref{fig:two}a) are summarized in Fig.~\ref{fig:three}. More precisely, in Fig.~\ref{fig:three}a) we plot our prediction for $Q$ as a function of the carrier density $n$, for a fixed value of the illumination wavelength $\lambda = 10.6~\mu{\rm m}$ (corresponding to a mid-infrared photon energy $\omega=110~{\rm meV}$), 
and at a temperature $T=300~{\rm K}$. As a comparison, we also plot the ``inverse damping ratio'' $Q_{\rm tr}$ calculated by replacing the HPP lifetime $\tau(q_1,\omega)$ with the transport time $\tau_{\rm tr}$ in the denominator of Eq.~(\ref{eq:Q_final_expression}). Note that $Q_{\rm tr} \gg Q$ and that $Q$ has a rather weak density dependence. 
We conclude that the HPP inverse damping ratio at $10.6~{\rm \mu m}$ and at room temperature falls in the range $50$-$70$ in hBN/G/hBN stacks with carrier mobilities $\lesssim 35.000~{\rm cm}^2/({\rm V} {\rm s})$---see panel b) in Fig.~\ref{fig:three}. Interestingly, we notice that the density dependence of the mobility $\mu$ and that of the HPP mode inverse damping ratio $Q$ are not correlated: at large densities $\mu$ decreases, while $Q$ shows a slight increase. The mobility $\mu$ is, of course, correlated with the unphysical construct $Q_{\rm tr}$. Note also that, for typical carrier densities, the ratio $Q/Q_{\rm tr}$ is always significantly smaller than one---see panel Fig.~\ref{fig:three}c).

In Fig.~\ref{fig:three}d) we illustrate the dependence of $Q$ on illumination frequency, for a fixed value of the carrier density $n=7.2\times 10^{12}~{\rm cm}^{-2}$. Gaps in these curves occur when the illumination energy falls in the reststrahlen bands, where the collective HPP mode ceases to exist. 
As expected, the inverse damping ratio is strongly suppressed when the illumination energy approaches the reststrahlen bands. This results in a non-monotonic 
behavior of $Q$ in the region between the two reststrahlen bands.

\section{Scattering of HPP modes against hBN optical phonons} 
\label{sect:optical_substrate}
In this Section we calculate the inverse damping ratio of the HPP mode displayed in Fig.~\ref{fig:two}a) by taking into account scattering against hBN optical phonons. According to Eq.~(\ref{eq:lifetime_final_T}), two ingredients are necessary to calculate the lifetime of the HPP mode, namely the phonon frequencies and the electron-phonon coupling. A detailed derivation of these quantities is given in Appendix~\ref{sect:effective_frolich_H}. Here we briefly summarize the main results.

When the 3D phonon momentum ${\bm q}$ is either parallel or perpendicular to the ${\hat {\bm z}}$ axis, the phonon frequency coincides with either $\omega^{\rm L}_z$ or $\omega^{\rm L}_x$. However, if ${\bm q}$ is along any other direction the two modes are mixed, and two ``extraordinary phonons''~\cite{Lee_prb_1997} can be excited. These modes are neither longitudinal nor transverse. In Appendix~\ref{sect:effective_frolich_H} we show that the extraordinary phonon frequencies $\omega_{{\rm ph},\nu}({\bm q})$ ($\nu=1,2$) of bulk hBN can be found by solving the equation
\begin{equation}
q_\perp^2 \varepsilon_x \big(\omega_{{\rm ph},\nu}({\bm q})\big) + q_\parallel^2 \varepsilon_z \big(\omega_{{\rm ph},\nu}({\bm q})\big) = 0
~.
\end{equation}
Analytical expression for $\omega_{{\rm ph},\nu}({\bm q})$ are available but are rather complicated and will not be reported here. The electron-phonon interaction is given by
\begin{equation}\label{eq:eopphinteraction}
u_{{\bm q},\nu}^{\rm op} = \sqrt{\frac{4\pi e^2}{ \big[ q_\parallel^2 \partial_\omega \varepsilon_z(\omega) + q_\perp^2 \partial_\omega \varepsilon_x(\omega) \big]_{\omega=\omega_{{\rm ph}, \nu}({\bm q})} } }
~.
\end{equation}
The derivation of Eq.~(\ref{eq:eopphinteraction}) is also given in Appendix~\ref{sect:effective_frolich_H}.

Although the expressions of $\omega_{{\rm ph},\nu}({\bm q})$ and $u_{{\bm q},\nu}$ are rather cumbersome, we can greatly simplify the calculation by noting the following crucial identity (see Appendix~\ref{sect:effective_frolich_H} for a derivation),
\begin{equation} \label{eq:Frohlich_reduction_formula_Im_main}
\int_{-\infty}^{\infty} \frac{dq_\parallel}{2\pi} \sum_\nu |u_{{\bm q},\nu}^{\rm op}|^2 \Im m\big[{\cal D}^{({\rm ph})}_{\nu}(q,\omega) \big] = \Im m \big[V(q,\omega)\big]\,,
\end{equation}
where $V(q,\omega)$ is the phonon-mediated effective interaction between electrons in the graphene plane.

Using Eq.~(\ref{eq:Frohlich_reduction_formula_Im_main}) in Eq.~(\ref{eq:lifetime_final_T}) we finally get the following expression for the relaxation rate due to scattering against hBN optical phonons:
\begin{eqnarray} 
\frac{1}{\tau^{({\rm op})}(q,\omega)} &=& - \frac{\pi e^2\omega}{{\cal D}_0} \sum_{\alpha,\beta} \int \frac{d^2{\bm q}'}{(2\pi)^2} \int_{-\infty}^{\infty} \frac{d\omega'}{\pi} 
\Im m \big[V(q',\omega')\big]
\nonumber\\
&\times&
\big[n_{\rm B}(\omega') - n_{\rm B}(\omega'-\omega) \big]  \Gamma^{({\rm ph})}_\alpha({\bm q},{\bm q}',\omega,\omega') 
\nonumber\\
&\times&
\Gamma^{({\rm ph})}_\beta({\bm q},{\bm q}',\omega,\omega') 
\Im m\chi_{j_\alpha j_\beta}^{(0)} ({\bm q}+{\bm q}',\omega-\omega')
~.
\nonumber\\
\end{eqnarray}
We emphasize that the corresponding dc transport time $\tau^{({\rm op})}_{\rm tr} = \lim_{\omega\to 0} \tau^{({\rm op})}(q=0,\omega)$ {\it diverges}, since optical phonons require a finite energy to be excited. Even though the contribution of optical phonons to the dc transport scattering rate is negligible, this is not necessarily the case for the scattering rate $1/\tau(q,\omega)$ evaluated at finite $q$ and $\omega$, which is relevant for the HPP mode inverse damping ratio. An HPP mode can indeed decay by the simultaneous emission of an electron-hole pair and an optical phonon in the substrate.  We note that this process is physically distinct from the dissipation that arises from the finite lifetime of the optical phonons in the substrate---the $r(\omega)$ contribution.

\begin{figure}[t]
\begin{center}
\begin{tabular}{c}
\includegraphics[width=0.99\columnwidth]{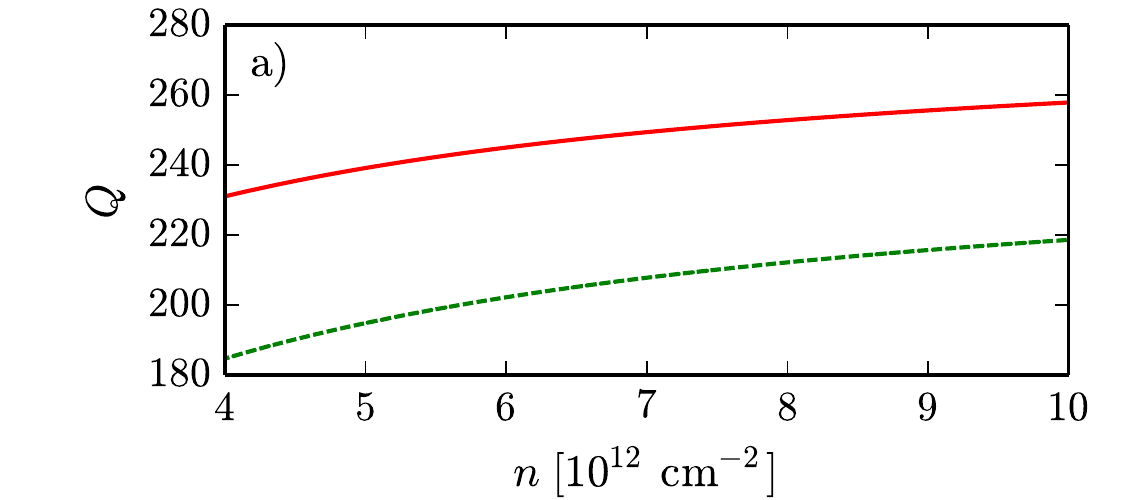}
\\
\includegraphics[width=0.99\columnwidth]{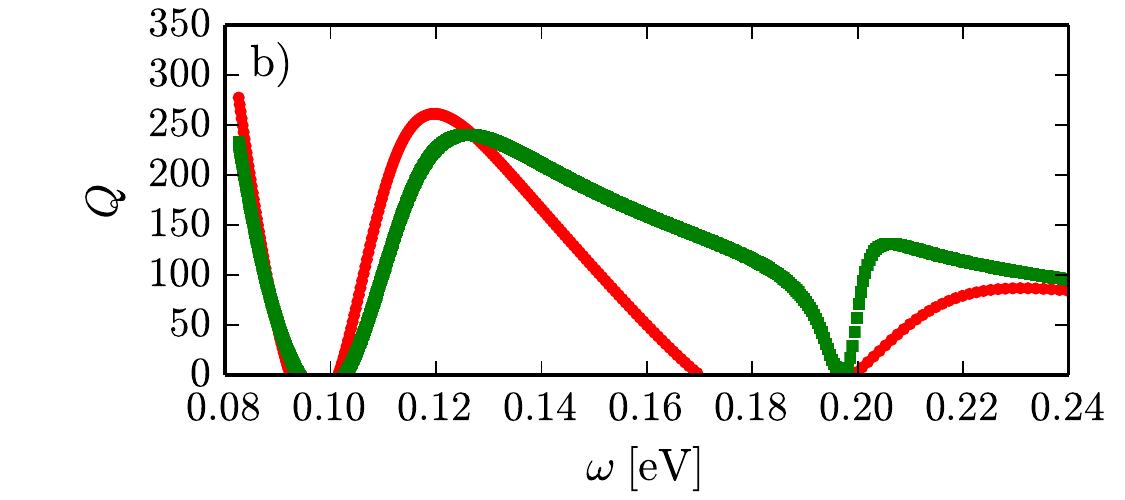}
\end{tabular}
\end{center}
\caption{(Color online) Panel a) the inverse damping ratio of graphene due to the scattering with optical phonons of the substrate (solid line). The curve is plotted as a function of density $n$ in units of $10^{12}~{\rm cm}^{-2}$ and for fixed illumination wavelength $\lambda = 10.6~\mu{\rm m}$ (corresponding to $\omega=110~{\rm meV}$) and temperature $T=0~{\rm K}$.  Note that the contribution $\Im m[\varepsilon_{\rm s}(\omega)]/\Re e[\varepsilon_{\rm s}(\omega)]$ completely dominates the  HPP damping rate. As a comparison we plot the inverse damping ratio obtained by removing the anisotropy of the substrate (dashed line).
Panel b) same as panel a) but plotted as a function of the illumination frequency (measured in ${\rm eV}$) and for fixed density $n = 7.2\times 10^{12}~{\rm cm}^{-2}$ and temperature $T=0~{\rm K}$.
\label{fig:four}}
\end{figure}

In Fig.~\ref{fig:four} we show our results for the HPP mode inverse damping ratio due to scattering against hBN optical phonons. In Fig.~\ref{fig:four}a) $Q$ is plotted as a function of density $n$ and for a fixed illumination wavelength (solid line). By comparing the numbers for $Q$ with those shown in Fig.~\ref{fig:one}, we conclude that the result is completely dominated by the extrinsic contribution $r(\omega)$ evaluated at $\omega = 110~{\rm meV}$. The dashed line in Fig.~\ref{fig:four}a) indicates the result that one obtains by artificially forcing $\varepsilon_x(\omega) \to \varepsilon_z(\omega)$. In Fig.~\ref{fig:four}b) we show the same quantities as in panel a) but this time plotted as functions of the illumination frequency $\omega$ and for a fixed carrier density $n = 7.2\times 10^{12}~{\rm cm}^{-2}$. In these plots the temperature has been fixed at the value $T=0~{\rm K}$: since the hBN optical phonon energy scales are very high, we do not expect any significant temperature dependence in the relevant range $0~{\rm K} \leq T \leq 300~{\rm K}$.

\section{Summary and conclusions}
\label{sect:summary}

We have presented a theory of hybrid plasmon-phonon modes in a graphene sheet encapsulated between two hexagonal Boron Nitride (hBN) semi-infinite slabs (hBN/G/hBN).  By using linear response theory and the random phase approximation, we have calculated the dispersion relation of 
hybrid plasmon-phonon modes that stem from the hybridization between graphene Dirac plasmons and hBN optical phonons. The uniaxial hyperbolic nature of hBN yields three plasmon-phonon branches separated by two reststrahlen bands.

We have carried out a detailed study of the inverse damping ratio of these plasmon-phonon modes. We have considered two possible sources of scattering limiting their lifetime: scattering against graphene's acoustic phonons and hBN optical phonons. We have discovered that scattering against intrinsic acoustic phonons is the dominant limiting factor in hBN/G/hBN stacks and that it yields theoretical inverse damping ratios of hybrid plasmon-phonon modes that lie in the range $50$-$70$ at room temperature, with a weak dependence on carrier density and a strong dependence on illumination wavelength. While the current work focuses on room temperature for its relevance to applications, the inverse damping ratio is expected to increase when temperature is lowered and the scattering of electrons with intrinsic acoustic phonons is suppressed. Numerical calculations on the temperature dependence of the inverse damping ratio at a fixed carrier density and illumination frequency will be shown elsewhere~\cite{Principi_TBS}.

Our theoretical predictions indicate that hBN/G/hBN stacks can be very fruitfully used as a low-loss and gate tunable platform for plasmonics in the mid-infrared spectral range. We will show somewhere else~\cite{Koppens} that our results are in excellent quantitative agreement with recent s-SNOM measurements.

\acknowledgments
It is a great pleasure to thank Andrea Tomadin and Leonid Levitov for many useful discussions. This work was supported by the DOE grant DE-FG02-05ER46203 (A.P. and G.V.), a Research Board Grant at the University of Missouri (A.P. and G.V.), the E.U. through the Graphene Flagship (contract no. CNECT-ICT-604391) program (M.C., M.L., A.W., F.H.L.K., and M.P.), and the Italian Ministry of Education, University, and Research (MIUR) through the programs ``FIRB - Futuro in Ricerca 2010" - Project PLASMOGRAPH (Grant No. RBFR10M5BT) and ``Progetti Premiali 2012" - Project ABNANOTECH (M.P.). M.C. acknowledges also the support of MIUR-FIRB2012 - Project HybridNanoDev (Grant No. RBFR1236VV).

\appendix

\section{The effective electron-electron interaction from electrostatics}
\label{sect:ee_interaction_electrostatics}

In this Appendix we derive the effective electron-electron interaction in a graphene sheet embedded between two semi-infinite uniaxial dielectrics. 

We assume to have a graphene sheet lying on the $z=0$ plane and surrounded by two semi-infinite uniaxial dielectrics filling the half spaces $z<0$ and $z>0$. To determine the effective electron-electron interaction we solve the electrostatic problem 
\begin{equation} \label{eq:app_electrostatic_problem}
\big[ \varepsilon_z(\omega) \partial_z^2 + \varepsilon_x(\omega) {\bm \nabla}_{\bm r}^2 \big] \phi({\bm r},z) = -4\pi e n({\bm r}) \delta(z)
\end{equation}
for the 3D electrical potential $\phi({\bm r},z)$. Here ${\bm r}=(x,y)$ is a 2D vector in the $z=0$ plane, ${\bm \nabla}_{\bm r} = (\partial_x,\partial_y)$, and $n({\bm r})$ is the charge density on the graphene sheet. The in-plane $\varepsilon_x(\omega)$ and out-of-plane $\varepsilon_z(\omega)$ components of the hBN dielectric tensor have been defined earlier in Eq.~(\ref{eq:BN_dielectric}).

We solve Eq.~(\ref{eq:app_electrostatic_problem}) in the two half-spaces $z>0$ and $z<0$ with the {\it Ansatz} 
\begin{equation} \label{eq:app_ansatz_es}
\phi({\bm r},z) = \phi_{{\bm q}_\perp} \exp( i {\bm q}_\perp\cdot{\bm r} - q_{\parallel} |z|)~.
\end{equation}
Substituting Eq.~(\ref{eq:app_ansatz_es}) in Eq.~(\ref{eq:app_electrostatic_problem}) we find, for $z \neq 0$, 
\begin{equation}
\big[ \varepsilon_z(\omega) q_{\parallel}^2 - \varepsilon_x(\omega) q_\perp^2 \big] \phi_{{\bm q}_\perp} = 0~,
\end{equation}
which is solved by the choice
\begin{eqnarray} \label{eq:app_q_z_hyperbolic}
q_\parallel = q_\perp\sqrt{\frac{\varepsilon_x(\omega)}{\varepsilon_z(\omega)}}~.
\end{eqnarray}
Note that $q_\parallel$ becomes imaginary when the product $\varepsilon_z(\omega) \varepsilon_x(\omega)<0$. In this case the {\it Ansatz} (\ref{eq:app_ansatz_es}) describes electromagnetic waves that are not bounded to the graphene sheet and that propagate in all the three spatial directions. 

The quantity $\phi_{{\bm q}_\perp}$ is determined by matching the $z<0$ and $z>0$ solutions at $z=0$ with the ``metallic'' boundary condition imposed by the presence of graphene, i.e.
\begin{equation}
\varepsilon_z(\omega)[\partial_z \phi({\bm r},z)|_{z=0^+} - \partial_z \phi({\bm r},z)|_{z=0^-}]  = - 4\pi e~,
\end{equation}
where $-e$ is the electron charge.

The requested 2D effective electron-electron interaction $V(q_\perp,\omega)$ is simply given by
\begin{equation} \label{eq:app_effective_interaction}
V(q_\perp,\omega) \equiv -e\phi_{{\bm q}_\perp} = \frac{2\pi e^2}{\varepsilon_{\rm s}(\omega) q_\perp}
\end{equation}
with
\begin{equation} \label{eq:barepsilon_sandwich}
\varepsilon_{\rm s}(\omega) = \sqrt{ \varepsilon_x(\omega) \varepsilon_z(\omega)}~.
\end{equation}

\section{The response function at finite temperature}
\label{app:response_function_finite_T}
We start from the definition of the finite-temperature (Matsubara) linear response function
\begin{equation} \label{eq:chi_AB_tau_def}
\chi^{({\cal T})}_{AB}(\tau) = - \big\langle {\cal T}({\hat A}(\tau) {\hat B})\big\rangle~,
\end{equation}
where $\langle\ldots\rangle$ stands for the average over the finite-temperature density matrix and ${\cal T}$ indicates the imaginary-time ordering operator. We assume that ${\hat A}$ and ${\hat B}$ are bosonic operators built with an even number of fermionic operators. In the frequency representation $\chi^{({\cal T})}_{AB}$ reads as follows:
\begin{equation} \label{eq:chi_AB_omega_def}
\chi_{AB}(i\omega_m) = \int_0^\beta d\tau e^{i\omega_m \tau} \chi^{({\cal T})}_{AB}(\tau)~,
\end{equation}
where $\omega_m =2\pi m/\beta$ is a bosonic Matsubara frequency and $\beta=(k_{\rm B} T)^{-1}$ is the inverse temperature. More specifically, ${\hat A} = ({\hat a}^\dagger_{{\bm q}',\nu} + {\hat a}_{-{\bm q}',\nu}) {\hat \Upsilon}_{{\bm q},{\bm q}'_\perp}$ and ${\hat B} = ({\hat a}^\dagger_{-{\bm q}'',\nu'} + {\hat a}_{{\bm q}'',\nu'}) {\hat \Upsilon}_{-{\bm q},-{\bm q}''_\perp}$ where the operator ${\hat \Upsilon}_{{\bm q},{\bm q}'}$ has been introduced earlier in Eq.~(\ref{eq:Upsilon_approx}).

 Since $\tau>0$ in Eq.~(\ref{eq:chi_AB_omega_def}), we can remove the imaginary-time ordering operator ${\cal T}$ on the right-hand side of Eq.~(\ref{eq:chi_AB_tau_def}). Furthermore, we approximate
\begin{eqnarray} \label{chi_tau_approx}
{\cal M} &=& \big\langle \big[{\hat a}^\dagger_{{\bm q}',\nu}(\tau) + {\hat a}_{-{\bm q}',\nu}(\tau)\big] {\hat \Upsilon}_{{\bm q},{\bm q}'_\perp}(\tau) 
\nonumber\\
&\times&
({\hat a}^\dagger_{-{\bm q}'',\nu'} + {\hat a}_{{\bm q}'',\nu'}) {\hat \Upsilon}_{-{\bm q},-{\bm q}''_\perp} \big\rangle
\nonumber\\
&\simeq&
\big\langle\! \big[{\hat a}^\dagger_{{\bm q}',\nu}(\tau) + {\hat a}_{-{\bm q}',\nu}(\tau)\big] ({\hat a}^\dagger_{-{\bm q}'',\nu'} + {\hat a}_{{\bm q}'',\nu'}) \!\big\rangle 
\nonumber\\
&\times&
\big\langle\! {\hat \Upsilon}_{{\bm q},{\bm q}'_\perp}(\tau)  {\hat \Upsilon}_{-{\bm q},-{\bm q}''_\perp} \!\big\rangle
\nonumber\\
&=& 
\delta_{{\bm q}',{\bm q}''} \delta_{\nu,\nu'} {\cal D}^{({\rm ph})}_\nu({\bm q}',\tau) \chi_{\Upsilon_{{\bm q},{\bm q'}_\perp} \Upsilon_{-{\bm q},-{\bm q}'_\perp}}^{\cal T}(\tau) 
~,
\end{eqnarray}
where we retained only the Wick pairings that are dominant in the large-$N_{\rm f}$ limit.

Using the last line of Eq.~(\ref{chi_tau_approx}) in Eq.~(\ref{eq:chi_AB_omega_def}) and the identity
\begin{equation}
\int_0^\beta d\tau e^{i\omega_n \tau} = \beta \delta_{n,0}
\end{equation}
which is valid for any bosonic Matsubara frequency $\omega_n$, we get
\begin{eqnarray} \label{eq:chi_jnjn_omega}
\chi_{j_{1,x} j_{1,x}}({\bm q}, i\omega_m) &=& -\frac{1}{\beta} \sum_{\omega_n, {\bm q}',\nu} u_{{\bm q}',\nu}^2
{\cal D}_\nu^{({\rm ph})}({\bm q}',i\omega_n) 
\nonumber\\
&\times&
\chi_{\Upsilon_{{\bm q},{\bm q'}_\perp} \Upsilon_{-{\bm q},-{\bm q}'_\perp}}(i\omega_m - i\omega_n)
\nonumber\\
&\equiv&
-\frac{1}{\beta} \sum_{\omega_n} f(i\omega_n,i\omega_m-i\omega_n)
~.
\end{eqnarray}

The Matsubara sum in the last line of Eq.~(\ref{eq:chi_jnjn_omega}) can be transformed into an integral over a contour in the complex plane that encircles the poles of  the Bose-Einstein occupation factor $n_{\rm B}(z) = (e^{\beta z} - 1)^{-1}$. In doing so, we exclude the branch cuts of $f(z,i\omega_m-z)$ which occur for $\Im m(z) = 0, \omega_m$. After the analytical continuation to real frequencies we find
\begin{eqnarray} \label{chi_jn_jn_causal_2}
&& \!\!\!\!\!\!\!\!
\chi_{j_{1,x} j_{1,x}}({\bm q},\omega) = \int_{-\infty}^{+\infty} \frac{d\omega'}{2\pi i} \Big\{
\big[n_{\rm B}(\omega') - n_{\rm B}(\omega'-\omega) \big] 
\nonumber\\
&\times&
\big[ f_{++}(\omega',\omega-\omega') - f_{--}(\omega',\omega-\omega') \big]
\nonumber\\
&+& n_{\rm B}(\omega') \big[ f_{+-}(\omega',\omega-\omega') - f_{-+}(\omega',\omega-\omega') \big]
\Big\}
~,
\nonumber\\
\end{eqnarray}
where $f_{\lambda\lambda'}(\omega,\omega') = f(\omega+i\lambda\eta,\omega'+i\lambda'\eta)$ and $\eta = 0^+$. Note that the term in square bracket in the last line of Eq.~(\ref{chi_jn_jn_causal_2}) is purely imaginary. Since it is multiplied by the imaginary unit, its contribution to the integral is purely real. Taking the imaginary part of Eq.~(\ref{chi_jn_jn_causal_2}) we finally find
\begin{eqnarray} \label{Im_chi_jn_jn_causal}
\Im m\chi_{j_{1,x} j_{1,x}}({\bm q},\omega) &=& -\sum_{{\bm q}',\nu} u_{{\bm q}',\nu}^2 \int_{-\infty}^{+\infty} \frac{d\omega'}{\pi}
\Im m {\cal D}_\nu^{({\rm ph})}({\bm q}',\omega') 
\nonumber\\
&\times&
\big[n_{\rm B}(\omega') - n_{\rm B}(\omega'-\omega) \big] 
\nonumber\\
&\times&
\Im m\chi_{\Upsilon_{{\bm q},{\bm q'}_\perp} \Upsilon_{-{\bm q},-{\bm q}'_\perp}}(\omega - \omega')
~.
\end{eqnarray}
In the limit $T\to 0$, and for $\omega>0$, Eq.~(\ref{Im_chi_jn_jn_causal}) becomes
\begin{eqnarray} \label{Im_chi_jn_jn_causal_T0}
\Im m\chi_{j_{1,x} j_{1,x}}({\bm q},\omega) &=& -\sum_{{\bm q}',\nu} u_{{\bm q}',\nu}^2 \int_{0}^{\omega} \frac{d\omega'}{\pi}
\Im m {\cal D}_\nu^{({\rm ph})}({\bm q}',\omega') 
\nonumber\\
&\times&
\Im m\chi_{\Upsilon_{{\bm q},{\bm q'}_\perp} \Upsilon_{-{\bm q},-{\bm q}'_\perp}}(\omega - \omega')
~.
\end{eqnarray}

\section{The dc transport time due to scattering of electrons against graphene's acoustic phonons} 
\label{app:transport_time}
In this Appendix we show that, in the limit $q=0$ and $\omega \to 0$, Eq.~(\ref{eq:lifetime_final_T}) reproduces the dc transport time $\tau_{\rm tr}$ for scattering of electrons against graphene's acoustic phonons as found e.g.~in Ref.~\onlinecite{Hwang_prb_2008}.

In the limit above, the matrix element $\Gamma_\alpha({\bm q},{\bm q}',\omega,\omega')$ defined in Eq.~(\ref{eq:Upsilon_approx}) reduces to
\begin{equation} \label{eq:app_matr_elem_transport_limit}
\Gamma_\alpha({\bm 0},{\bm q}',\omega,\omega') = \frac{v_{\rm F} q'_x q'_\alpha}{ k_{\rm F} \omega (\omega-\omega')}~.
\end{equation}
Substituting Eq.~(\ref{eq:app_matr_elem_transport_limit}) in Eq.~(\ref{eq:lifetime_final_T}) we therefore get
\begin{eqnarray} \label{eq:transport_time_1}
\frac{1}{\tau_{\rm tr}} &=& \frac{\pi e^2}{{\cal D}_0} \sum_{{\bm q}',\alpha,\beta} |u^{({\rm ac})}_{{\bm q}'}|^2 \int_{-\infty}^{\infty} \frac{d\omega'}{\pi} \partial_{\omega'} n_{\rm B}(\omega') \frac{v_{\rm F}^2 q_x'^2 q'_\alpha q'_\beta}{ k_{\rm F}^2 (\omega-\omega')^2} 
\nonumber\\
&\times&
\Im m~[{\cal D}^{({\rm ph})}(q',\omega')]\Im m~[\chi_{j_\alpha j_\beta}^{(0)} ({\bm q}',\omega')]
~.
\end{eqnarray}
We now recall~\cite{Giuliani_and_Vignale} that, in a homogeneous and isotropic electron liquid, the current-current linear response tensor can be decomposed in the following manner:
\begin{eqnarray} \label{eq:current_decomposition}
\chi_{j_\alpha j_\beta}^{(0)} ({\bm q},\omega) &=& \frac{q_\alpha q_\beta}{q^2} \chi_{\rm L}^{(0)} (q,\omega) \nonumber\\
&+&\left(\delta_{\alpha\beta}-\frac{q_\alpha q_\beta}{q^2}\right) \chi_{\rm T}^{(0)} (q,\omega)~,
\end{eqnarray}
where $\chi_{\rm L}^{(0)} (q,\omega)$ and $\chi_{\rm T}^{(0)} (q,\omega)$ are the so-called~\cite{Giuliani_and_Vignale} longitudinal and transverse current-current response functions, respectively. Using Eq.~(\ref{eq:current_decomposition}) we can write Eq.~(\ref{eq:transport_time_1}) in the following form:
\begin{eqnarray} \label{eq:transport_time_2}
\frac{1}{\tau_{\rm tr}} &=& 
\frac{\pi e^2}{{\cal D}_0} \frac{v_{\rm F}^2}{2 k_{\rm F}^2}  \sum_{{\bm q}'} q'^2 |u^{({\rm ac})}_{{\bm q}'}|^2 \int_{-\infty}^{\infty} \frac{d\omega'}{\pi} \partial_{\omega'} n_{\rm B}(\omega') 
\nonumber\\
&\times& 
\Im m~[{\cal D}^{({\rm ph})} (q',\omega')]\Im m~[\chi_{nn}^{(0)} (q',\omega')]~.
\end{eqnarray}
Here we used Eq.~(\ref{eq:continuity_equation}) to express the longitudinal current-current response function in terms of the density-density response function.

Using Eq.~(\ref{eq:acoustic_phonon_interaction}) in Eq.~(\ref{eq:transport_time_2}) and the imaginary part of the phonon propagator in the absence of phonon damping, i.e.
\begin{eqnarray} \label{eq:phonon_propagator_nodamping}
\Im m~[{\cal D}^{({\rm ph})}(q',\omega')] &=& -\pi[\delta(\omega'-\omega_{\rm ph}(q')) \nonumber \\
&-& \delta(\omega'+\omega_{\rm ph}(q'))]~,
\end{eqnarray}
we find
\begin{eqnarray} \label{eq:transport_time_3}
\frac{1}{\tau_{\rm tr}} &=& 
- 2 \frac{\pi e^2}{{\cal D}_0} \frac{v_{\rm F}^2}{2 k_{\rm F}^2}  \int \frac{d^2 {\bm q}'}{(2\pi)^2} q'^2 \frac{D^2 q'^2}{2 \rho \omega_{\rm ph}(q')} \partial_{\omega'} n_{\rm B}(\omega')\Big|_{\omega_{\rm ph}(q')}
\nonumber\\
&\times&\Im m[\chi_{nn}^{(0)}(q',\omega_{\rm ph}(q'))]~.
\end{eqnarray}
Note that the two terms on the right-hand side of Eq.~(\ref{eq:phonon_propagator_nodamping}) give identical contributions to the integral in Eq.~(\ref{eq:transport_time_2}): this explains the factor two on the right-hand side of Eq.~(\ref{eq:transport_time_3}).

We now specialize Eq.~(\ref{eq:transport_time_3}) to the case in which the phonon frequency is much smaller than both  temperature and Fermi energy. In this case we use the following approximate expressions:
\begin{equation}
\partial_{\omega'} n_{\rm B}(\omega')\Big|_{\omega_{\rm ph}(q')} \to -\frac{k_{\rm B} T}{\omega_{\rm ph}^2(q')}
~,
\end{equation}
and 
\begin{eqnarray}
\Im m~[\chi_{nn}^{(0)} (q',\omega')] &\to& -\Theta(2 k_{\rm F} - q')\frac{\omega'}{v_{\rm F} q'} \frac{N_{\rm f} k_{\rm F}}{2\pi v_{\rm F}} \nonumber \\
&\times &\sqrt{1-\frac{q'^2}{4k_{\rm F}^2}}~.
\end{eqnarray}
After some straightforward algebra Eq.~(\ref{eq:transport_time_3}) gives
\begin{equation} \label{eq:transport_time_4}
\frac{1}{\tau_{\rm tr}} =\frac{\varepsilon_{\rm F}}{4 v_{\rm F}^2} \frac{D^2}{\rho v_{\rm ph}^2} k_{\rm B} T~,
\end{equation}
which coincides with the result reported in Ref.~\onlinecite{Hwang_prb_2008}.

\section{The derivation of the Fr\"ohlich Hamiltonian}
\label{sect:effective_frolich_H}
In this Appendix we derive the interaction Hamiltonian between electrons in graphene and optical phonons in hBN. In what follows we shorten our notation by setting $\omega_\ell \equiv \omega_\ell^{\rm T}$. 

Following Ref.~\onlinecite{Lee_prb_1997} we start from the following equations
\begin{eqnarray} \label{eq:stroscio_1}
\partial_t^2 u_{\ell} ({\bm r}, t) &=& -\omega^2_{\ell} u_\ell({\bm r},t) + \omega_\ell \sqrt{\frac{\epsilon_{\ell,0} - \epsilon_{\ell,\infty}}{4\pi {\bar m} n_{\rm c}}} E_\ell({\bm r},t)
~,
\nonumber\\
P_\ell({\bm r}, t) &=& \sqrt{\frac{{\bar m} n_{\rm c} (\epsilon_{\ell,0} - \epsilon_{\ell,\infty})}{4\pi}} \omega_{\ell} u_\ell({\bm r},t) 
\nonumber\\
&+&
\frac{\epsilon_{\ell,\infty} - 1}{4\pi} E_\ell({\bm r}, t)
~,
\end{eqnarray}
which describe the coupling between the electric field $E_\ell({\bm r},t)$ and lattice motion, which is encoded in the displacement field $u_\ell({\bm r},t)$ between the two atoms in the hBN unit cell. The latter induces a polarization $P_\ell({\bm r}, t)$, which adds to the electric field to produce the electric displacement $D_\ell({\bm r},t) = E_\ell({\bm r},t) + 4\pi P_\ell({\bm r}, t)$. Finally, ${\bar m}$ is the reduced mass and $n_{\rm c}$ the number of cells in the unit volume. 

Eqs.~(\ref{eq:stroscio_1}) are combined with the following ones
\begin{eqnarray} \label{eq:stroscio_2}
&& {\bm E}({\bm r},t) = -{\bm \nabla} \Phi({\bm r},t)~,
\nonumber\\
&& {\bm D}({\bm r},t) = \varepsilon_x (\omega) E_\perp({\bm r},t) {\hat {\bm r}} + \varepsilon_z(\omega) E_\parallel({\bm r},t) {\hat {\bm z}}~,
\nonumber\\
&& {\bm \nabla} \cdot{\bm D}({\bm r},t) = 0~,
\end{eqnarray}
to solve the electrostatic problem. Here $\Phi({\bm r},t)$ is the electrostatic potential.

Fourier transforming Eqs.~(\ref{eq:stroscio_1}) and~(\ref{eq:stroscio_2}) with respect to space and time we find
\begin{subequations} \label{eq:stroscio_FT}
\begin{align}
& u_{\ell} ({\bm q}, \omega) = \frac{\omega_\ell}{\omega_\ell^2 - \omega^2} \sqrt{\frac{\epsilon_{\ell,0} - \epsilon_{\ell,\infty}}{4\pi {\bar m} n_{\rm c}}} E_\ell({\bm q},\omega)
~,
\label{eq:stroscio_FT_1}
\\
& P_\ell({\bm q}, \omega) = \frac{1}{4\pi} \big[\varepsilon_\ell(\omega) - 1\big] E_\ell({\bm q}, \omega)
~,
\label{eq:stroscio_FT_2}
\\
& E_\ell({\bm q},\omega) = -i q_\ell \Phi({\bm q},\omega)
~,
\label{eq:stroscio_FT_3}
\\
& q_\perp \varepsilon_x(\omega) E_\perp({\bm q},\omega) + q_\parallel \varepsilon_z(\omega) E_\parallel({\bm q},\omega) = 0
~.
\label{eq:stroscio_FT_4}
\end{align}
\end{subequations}
Eq.~(\ref{eq:stroscio_FT_3}) implies that the electric field is purely longitudinal, {\it i.e.} it is parallel to ${\bm q}$. Combining Eqs.~(\ref{eq:stroscio_FT_3}) and~(\ref{eq:stroscio_FT_4}) we get the following necessary condition
\begin{equation}
q_\perp^2 \varepsilon_x (\omega) + q_\parallel^2 \varepsilon_z (\omega) = 0
\end{equation}
to make sure that ${\bm \nabla}\cdot {\bm D}$ vanishes. The solutions $\{\omega_{{\rm ph},\nu}({\bm q}),~\nu=1,2\}$ of the previous equation are the so-called ``extraordinary'' phonon frequencies~\cite{Lee_prb_1997}. These modes are neither longitudinal nor transverse, except in the limiting cases of ${\bm q}$ oriented parallel or perpendicular to the ${\hat {\bm z}}$ axis. Indeed, Eq.~(\ref{eq:stroscio_FT_1}) implies that in general the lattice displacement is neither parallel nor perpendicular to the electric field and therefore to ${\bm q}$. Recall that ${\bm E}$ and ${\bm q}$ are always parallel.

We now define $u_{\ell,\nu}({\bm q},\omega)$ as the lattice displacement associated to the normal mode characterized by the frequency $\omega_{{\rm ph},\nu}({\bm q})$. Since normal modes are orthogonal to each other, combining Eqs.~(\ref{eq:stroscio_FT_1}) and~(\ref{eq:stroscio_FT_3}) we get the electrostatic potential
\begin{eqnarray} \label{eq:phi_u}
\Phi({\bm q}) &\equiv&
\sum_\nu \Phi\big({\bm q}, \omega_{{\rm ph},\nu}({\bm q})\big)
\nonumber\\
&=&
\frac{i}{q} \sum_{\ell,\nu} \sqrt{\frac{4\pi {\bar m} n_{\rm c}}{\epsilon_{\ell,0} - \epsilon_{\ell,\infty}}} \frac{\omega_\ell^2 - \omega_{{\rm ph},\nu}^2({\bm q})}{\omega_\ell} \frac{q_\ell u_{\ell,\nu}({\bm q},\omega)}{q}
~.
\nonumber\\
\end{eqnarray}
Once this quantity is quantized, it completely determines the electron-phonon interaction Hamiltonian, which is given by ${\hat {\cal H}}_{\rm e-ph} = e\sum_{\bm q} {\hat n}_{-{\bm q}} \Phi({\bm q})$. In what follows we proceed to quantize Eq.~(\ref{eq:phi_u}).

Eq.~(\ref{eq:stroscio_FT_1}) implies that the displacement can be written as
\begin{eqnarray} \label{eq:u_ell_components}
u_{\ell,\nu} ({\bm q}, \omega) &=& \left[ \sum_j \frac{\omega_j^2 (\epsilon_{j,0} - \epsilon_{j,\infty})}{[\omega_j^2 - \omega_{{\rm ph},\nu}^2({\bm q})]^2} \frac{q_j^2}{q^2} \right]^{-1/2} 
\nonumber\\
&\times&
\frac{\omega_\ell \sqrt{\epsilon_{\ell,0} - \epsilon_{\ell,\infty}}}{\omega_\ell^2 - \omega_{{\rm ph},\nu}^2({\bm q})} \frac{q_\ell}{q} u_{\nu}({\bm q})
~,
\end{eqnarray}
where $u_{\nu}({\bm q})$ is the modulus of the displacement vector ${\bm u}_\nu({\bm q})$, {\it i.e.} $u_{\nu}({\bm q}) = \sqrt{u_{\parallel,\nu}^2({\bm q}) + u_{\perp,\nu}^2({\bm q})}$. The prefactor of $u_{\nu}({\bm q})$ on the right-hand side of Eq.~(\ref{eq:u_ell_components}) is uniquely determined by the angle between the vectors ${\bm u}_\nu({\bm q})$ and ${\bm q}$, and can easily be derived from Eq.~(\ref{eq:stroscio_FT_1}). Quantizing the lattice displacement in the usual manner, i.e.~$u_{\nu}({\bm q}) = -i( a_{{\bm q},\nu} + a^\dagger_{-{\bm q},\nu})/\sqrt{2 n_{\rm c}{\bar m}V\omega_{{\rm ph},\nu}({\bm q})}$, with $V$ the volume of the system, we obtain
\begin{eqnarray} \label{eq:u_ell_def}
u_{\ell, \nu} ({\bm q}, \omega) &=& \frac{i (n_{\rm c} V {\bar m})^{-1/2}}{\sqrt{2  \omega_{{\rm ph},\nu}({\bm q})}} 
\left[ \sum_j \frac{\omega_j^2 (\epsilon_{j,0} - \epsilon_{j,\infty})}{[\omega_j^2 - \omega_{{\rm ph},\nu}^2({\bm q})]^2} \frac{q_j^2}{q^2} \right]^{-1/2} 
\nonumber\\
&\times&
\frac{\omega_\ell \sqrt{\epsilon_{\ell,0} - \epsilon_{\ell,\infty}}}{\omega_{{\rm ph},\nu}^2({\bm q}) - \omega_\ell^2} \frac{q_\ell}{q} \big( a_{{\bm q},\nu} + a^\dagger_{-{\bm q},\nu} \big)
~.
\end{eqnarray}
With the help of Eq.~(\ref{eq:BN_dielectric}) it is easy to show that the term in square brackets on the right-hand side of Eq.~(\ref{eq:u_ell_def}) gives
\begin{eqnarray} \label{eq:square_bracket_derivative_epsilon}
&& \!\!\!\!\!\!\!\!
{\cal S} = \sum_j \frac{\omega_j^2 (\epsilon_{j,0} - \epsilon_{j,\infty})}{[\omega_j^2 - \omega_{{\rm ph},\nu}^2({\bm q})]^2} \frac{q_j^2}{q^2} 
\nonumber\\
&=&
\frac{ \big[ q_\parallel^2 \partial_\omega \varepsilon_z(\omega) + q_\perp^2 \partial_\omega \varepsilon_x(\omega) \big]_{\omega=\omega_{{\rm ph},\nu}({\bm q})}}{2 q^2 \omega_{{\rm ph},\nu}({\bm q})}
~.
\nonumber\\
\end{eqnarray}
When Eq.~(\ref{eq:square_bracket_derivative_epsilon}) is used in Eq.~(\ref{eq:u_ell_def}) we obtain
\begin{eqnarray} \label{eq:u_ell_def_2}
&& \!\!\!\!\!\!\!\!
u_{\ell, \nu} ({\bm q}, \omega) = -i \frac{1}{\sqrt{{\bar m} n_{\rm c} V} }
\frac{\omega_\ell \sqrt{\epsilon_{\ell,0} - \epsilon_{\ell,\infty}}}{\omega_\ell^2 - \omega_{{\rm ph},\nu}^2({\bm q})} 
\nonumber\\
&\times&
\frac{q_\ell \big( a_{{\bm q},\nu} + a^\dagger_{-{\bm q},\nu} \big)}{\big[ q_\parallel^2 \partial_\omega \varepsilon_z(\omega) + q_\perp^2 \partial_\omega \varepsilon_x(\omega) \big]_{\omega=\omega_{{\rm ph},\nu}({\bm q})}^{1/2}}~.
\nonumber\\
\end{eqnarray}
Combining Eqs.~(\ref{eq:phi_u}) and~(\ref{eq:u_ell_def_2}) we finally get
\begin{eqnarray} 
&& \!\!\!\!\!\!\!\!\!
\Phi({\bm q}) = \sum_{\nu} \sqrt{ \frac{4\pi V^{-1}}{ \big[ q_\parallel^2 \partial_\omega \varepsilon_z(\omega) + q_\perp^2 \partial_\omega \varepsilon_x(\omega) \big]_{\omega=\omega_{{\rm ph},\nu}({\bm q})} } }
\nonumber\\
&\times&
\big( a_{{\bm q},\nu} + a^\dagger_{-{\bm q},\nu} \big) 
~.
\nonumber\\
\end{eqnarray}
This in turn implies that the electron-phonon Hamiltonian is
\begin{equation} \label{eq:el_phon_hamilt}
{\hat {\cal H}}_{\rm e-ph} = \sum_{{\bm q},\nu} u_{{\bm q},\nu} {\hat n}_{-{\bm q}} (a_{{\bm q},\nu} + a^\dagger_{-{\bm q},\nu})
~,
\end{equation}
where
\begin{equation}
u_{{\bm q},\nu} = \sqrt{\frac{4\pi e^2 V^{-1}}{ \big[ q_\parallel^2 \partial_\omega \varepsilon_z(\omega) + q_\perp^2 \partial_\omega \varepsilon_x(\omega) \big]_{\omega=\omega_{{\rm ph},\nu}({\bm q})} } }
~.
\end{equation}
Let us briefly comment this equation. First of all, if the system is isotropic, {\it i.e.} $\varepsilon(\omega) \equiv \varepsilon_x (\omega) = \varepsilon_z (\omega)$, the frequencies of the normal modes become $\omega_{{\rm ph},\nu}({\bm q}) = \{\omega^{\rm L}, \omega^{\rm T} \}$ (we omit the direction index, since the two are equivalent). The second solution, corresponding to a purely transverse mode, should be disregarded (see below). Substituting $\omega_{{\rm ph},\nu}({\bm q})=\omega^{\rm L}$ one immediately recovers the usual Fr\"ohlich Hamiltonian.

Moreover, when ${\bm q}$ is parallel (perpendicular) to the ${\hat {\bm z}}$ axis $\omega_{{\rm ph},1}({\bm q})$ coincides with $\omega^{\rm L}_z$ ($\omega^{\rm L}_x$), while $\omega_{{\rm ph},2}({\bm q})$ is equal to $\omega^{\rm T}_x$ ($\omega^{\rm T}_z$). This in turn implies that the mode $\nu=1$ is purely longitudinal, while $\nu=2$ describes a transverse optical phonon. However, the denominator of Eq.~(\ref{eq:el_phon_hamilt}), {\it i.e.} $\big[ q_\parallel^2 \partial_\omega \varepsilon_z(\omega) + q_\perp^2 \partial_\omega \varepsilon_x(\omega) \big]_{\omega=\omega_{{\rm ph},\nu}({\bm q})}$, diverges when $\omega_{{\rm ph},\nu}({\bm q})$ coincides with one of the transverse frequencies, thus excluding the transverse modes from the electron-phonon interaction.

By adding the phonon-mediated electron-electron interactions, derived from Eq.~(\ref{eq:el_phon_hamilt}), to the bare Coulomb potential one finds the effective 3D electron-electron interactions
\begin{eqnarray} \label{eq:v_eff_Frohlich}
V_{\rm 3D}({\bm q},\omega) &=& \frac{4\pi e^2}{\varepsilon_{x,\infty} q_\perp^2 + \varepsilon_{z,\infty} q_\parallel^2} 
\nonumber\\
&+&
\sum_{\nu} \frac{4\pi e^2 {\cal D}^{({\rm ph})}_\nu({\bm q},\omega)}{ \big[ q_\parallel^2 \partial_\omega \varepsilon_z(\omega) + q_\perp^2 \partial_\omega \varepsilon_x(\omega) \big]_{\omega=\Omega_\nu({\bm q})} }~,
\nonumber\\
\end{eqnarray}
which turns out to be equal to
\begin{equation} \label{eq:v_eff_electrostatics}
V_{\rm 3D}({\bm q},\omega) = \frac{4\pi e^2}{\varepsilon_x(\omega) q_\perp^2 + \varepsilon_z(\omega) q_\parallel^2}
~.
\end{equation}
Note that it is possible to recover the effective 2D electron-electron interaction in Eq.~(\ref{eq:app_effective_interaction}) by integrating Eq.~(\ref{eq:v_eff_electrostatics}) over $q_\parallel$. Moreover, Eqs.~(\ref{eq:v_eff_Frohlich}) and~(\ref{eq:v_eff_electrostatics}) allows us to write
\begin{eqnarray} \label{eq:app_Stilde_def}
{\tilde {\cal S}} &\equiv&
\sum_\nu u_{{\bm q},\nu}^2 {\cal D}^{({\rm ph})}_{\nu}(q,\omega) 
\nonumber\\
&=&
\frac{4\pi e^2}{\varepsilon_x(\omega) q_\perp^2 + \varepsilon_z(\omega) q_\parallel^2} - \frac{4\pi e^2}{\varepsilon_{x,\infty} q_\perp^2 + \varepsilon_{z,\infty} q_\parallel^2}
~.
\nonumber\\
\end{eqnarray}
This equation is used in Sect.~\ref{sect:optical_substrate}, where the integral
\begin{equation} \label{eq:I_def}
{\cal I}({\bm q}_\perp,\omega) \equiv \int_{-\infty}^{\infty} \frac{dq_\parallel}{2\pi} \sum_\nu u_{{\bm q},\nu}^2 {\cal D}^{({\rm ph})}_{\nu}(q,\omega) 
\end{equation}
is rewritten as
\begin{equation} \label{eq:Frohlich_reduction_formula}
{\cal I}({\bm q}_\perp,\omega) = \frac{2\pi e^2}{q_\perp \sqrt{\varepsilon_x(\omega) \varepsilon_z(\omega)}} - \frac{2\pi e^2}{q_\perp \sqrt{\varepsilon_{x,\infty} \varepsilon_{z,\infty}}}
~,
\end{equation}
with the help of Eq.~(\ref{eq:app_Stilde_def}). Eq.~(\ref{eq:Frohlich_reduction_formula}) in particular implies that
\begin{eqnarray} \label{eq:Frohlich_reduction_formula_Im}
\Im m[{\cal I}({\bm q}_\perp,\omega)] &=& \Im m \left[\frac{2\pi e^2}{q_\perp \sqrt{\varepsilon_x(\omega) \varepsilon_z(\omega)}}\right]
\nonumber\\
&\equiv&
\Im m\big[V(q_\perp,\omega)\big]
~.
\end{eqnarray}

\begin{thebibliography}{77}
\bibitem{footnote_fitting}
	More precisely, the value of ${\widetilde D}$ is determined by constrainining 
	the theoretical mobility $\mu$, calculated on the basis of Eq.~(\ref{eq:mobility}), 
	to yield the value $\mu = 30.000~{\rm cm}^2/({\rm Vs})$ at a carrier density $n=4.8 \times 10^{12}~{\rm cm}^{-2}$ 
	and a temperature $T =300~{\rm K}$. This implies a deformation potential $D=48.3~{\rm eV}$. 
	Once the value of the deformation potential has been found, we can make predictions 
	for $\mu$ as a function of carrier density.
\end{thebibliography}
\end{document}